\documentclass{aa}  

\usepackage{graphicx}
\usepackage{txfonts}
\usepackage{comment}
\usepackage{adjustbox}

\usepackage{xspace}
\newcommand{\kms}{km~s$^{-1}$\xspace}
\newcommand{\ergs}{erg~s$^{-1}$\xspace}
\newcommand{\Msunyr}{M$_{\odot}$~yr$^{-1}$\xspace}

\newcommand{\MOKA}{\ensuremath{\mathrm{MOKA^{3D}}}}
\newcommand{ \ha} {H$\alpha$}

\newcommand{ \nii} {[N\,\textsc{ii}]}
\newcommand{ \feii} {[Fe\,\textsc{ii}]\xspace}
\newcommand{ \sii} {[S\,\textsc{ii}]\xspace}

\usepackage[breaklinks=true, colorlinks=true, citecolor=blue, linkcolor=blue,urlcolor=blue]{hyperref}

\usepackage{threeparttable}

\begin{document}

\title{JWST/NIRSpec insights into the circumnuclear region of Arp 220:\\ A detailed kinematic study}

\author{
Lorenzo Ulivi\inst{\ref{iUNITR}, \ref{iUNIFI}, \ref{iOAA}}\thanks{e-mail: lorenzo.ulivi1@unifi.it}
\and Michele Perna \inst{\ref{iCAB}}
\and Isabella Lamperti \inst{\ref{iCAB},\ref{iUNIFI}, \ref{iOAA}}
\and Santiago Arribas \inst{\ref{iCAB}}
\and Giovanni Cresci \inst{\ref{iOAA}}
\and Cosimo Marconcini \inst{\ref{iUNIFI}, \ref{iOAA}}
\and Bruno Rodr\'iguez~Del~Pino\inst{\ref{iCAB}}
\and Torsten B{\"o}ker\inst{\ref{iESOba}}
\and Andrew J. Bunker\inst{\ref{iOxf}}
\and Matteo Ceci \inst{\ref{iUNIFI}, \ref{iOAA}}
\and St\'ephane Charlot\inst{\ref{iSor}}
\and Francesco D'Eugenio\inst{\ref{iKav},\ref{iCav}}
\and Katja Fahrion\inst{\ref{iESAne}}
\and Roberto Maiolino\inst{\ref{iKav},\ref{iCav}, \ref{iUCL}}
\and Alessandro Marconi \inst{\ref{iUNIFI}, \ref{iOAA}}
\and Miguel Pereira-Santaella\inst{\ref{iIFF}}
%
}
\institute{
University of Trento, Via Sommarive 14, I-38123 Trento, Italy\label{iUNITR}
\and 
Università di Firenze, Dipartimento di Fisica e Astronomia, via G. Sansone 1, 50019 Sesto F.no, Firenze, Italy\label{iUNIFI}
\and 
INAF - Osservatorio Astrofisico di Arcetri, Largo E. Fermi 5, I-50125 Firenze, Italy\label{iOAA}
\and 
Centro de Astrobiolog\'ia (CAB), CSIC--INTA, Cra. de Ajalvir Km.~4, 28850 -- Torrej\'on de Ardoz, Madrid, Spain\label{iCAB}
\and
European Space Agency, c/o STScI, 3700 San Martin Drive, Baltimore, MD 21218, USA\label{iESOba}
\and
Department of Physics, University of Oxford, Denys Wilkinson Building, Keble Road, Oxford OX1 3RH, UK\label{iOxf}
\and 
Sorbonne Universit\'e, CNRS, UMR 7095, Institut d’Astrophysique de Paris, 98 bis bd Arago, 75014 Paris, France\label{iSor} 
\and 
Kavli Institute for Cosmology, University of Cambridge, Madingley Road, Cambridge, CB3 0HA, UK\label{iKav}
\and
Cavendish Laboratory - Astrophysics Group, University of Cambridge, 19 JJ Thomson Avenue, Cambridge, CB3 0HE, UK\label{iCav}
\and 
European Space Agency, European Space Research and Technology Centre, Keplerlaan 1, 2201 AZ Noordwijk, the Netherlands\label{iESAne}
\and
Department of Physics and Astronomy, University College London, Gower Street, London WC1E 6BT, UK\label{iUCL}
\and
Instituto de F\'isica Fundamental, CSIC, Calle Serrano 123, 28006 Madrid, Spain\label{iIFF}
}
    \titlerunning{ JWST/NIRSpec view of Arp 220}
    \authorrunning{Ulivi, L., et al.}

   \date{July 11, 2024}

 
  \abstract
   {The study of starburst and active galactic nuclei (AGN) feedback is crucial for understanding the regulation of star formation and the evolution of galaxies across cosmic time.
   Arp 220, the closest ultraluminous infrared galaxy (ULIRG), is in an advanced phase of a major merger with two distinct nuclei, and it shows evidence of multiphase (molecular, ionized, and neutral) and multiscale (from < 0.1 to > 5 kpc) outflows. Therefore, it represents an ideal system for investigating outflow mechanisms and feedback phenomena in detail.
    Using new JWST NIRSpec IFU observations, we investigated the spatially resolved gaseous (in both ionized and hot molecular phases) and stellar kinematics in the innermost 1 kpc. We decoupled the different gas kinematic components through multi-Gaussian fitting, 
     identifying two multiphase outflows, each associated with one nucleus, 
    with velocities up to $\sim 1000$~\kms. We also resolved 
    two counter-rotating discs around each nucleus embedded in a larger-scale rotational disk.  
    We compute the total (including ionized, cold, and hot molecular) outflow mass ($\approx 10^7$~M$_\odot$), the mass rate ($\approx 15$~\Msunyr), and the energetics ($\dot E_{out}\approx 10^{42}$~\ergs) for each nucleus,
    and we found that the ionized and hot molecular outflowing gas contribute around 2-30\% of the total mass and the energy of the outflows, as inferred from the combination of multiwavelength information. We discuss the possible origin of the outflows, finding no compelling evidence to prefer a starburst- or AGN-driven scenario.
    Regardless of their nature, outflows in Arp~220 propagate in multiple directions from parsec to kiloparsec scales, potentially impacting a significant portion of the host galaxy. This contrasts with isolated systems where outflows typically 
    follow a more collimated path or are limited to the central region of the galaxy and hence do not affect the interstellar medium throughout the entire galaxy. This study highlights the importance of investigating merging systems with multiwavelength facilities, including JWST/NIRSpec IFU, to obtain a comprehensive understanding of feedback mechanisms in galaxy evolution.

   
   }



    \maketitle
%

\section{Introduction}

\begin{figure*}[t!]
    \centering
    \includegraphics[width = 1 \textwidth]{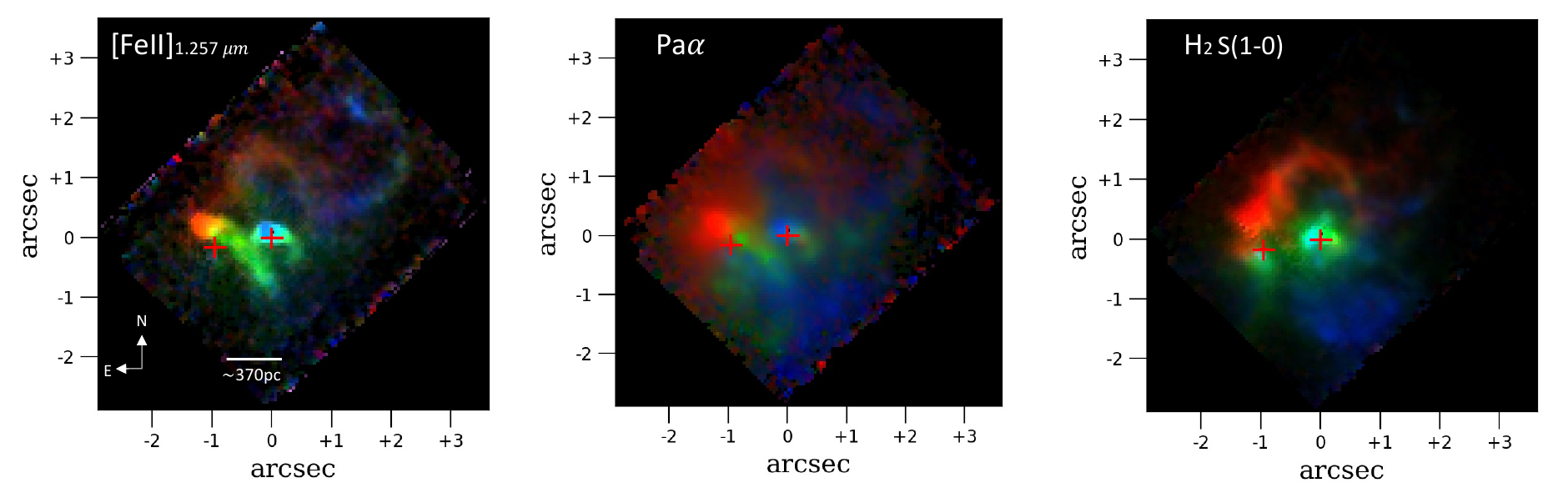}
    
    \caption{Three-color emission line images of Arp 220. The emission line maps are obtained integrating the continuum-subtracted data-cube around  \feii~1.257 $\mu m$ (left), Pa$\alpha$ (center), and H$_2$~1--0~S(1) 2.122 $\mu$m (right) in the velocity ranges $-400$ \kms < v < $-200$ \kms (blue), $-200$ \kms < v < 200 \kms (green) and 200 \kms < v < 400 \kms (red), with respect to the median value between $z_E^*$ and $z_W^*$, $z_m = 0.0181$. The red crosses mark the position of the two nuclei.}
    \label{RGB_Lines}
\end{figure*}


Ultraluminous infrared galaxies (ULIRGs) with infrared (IR) luminosities L$_{IR}$ > $10^{12}$ L$_{\odot}$ are among the most luminous objects in the local Universe. Their extreme IR luminosity is due to the combination of high star formation rate (SFR) and active galactic nuclei (AGN) activity, coupled with a high dust content, which intercepts the UV photons emitted from either hot young stars or AGN and reemits them at longer wavelengths (\citealt{soifer_1984,Sanders1996,Lonsdale2006, U2012} and references therein).

In the starburst-quasar evolutionary scenario (e.g., \citealt{Sanders1988, Hopkins2008}), ULIRGs are thought to represent a rapidly growing phase of massive galaxies. Within this framework, the merger of gas-rich galaxies triggers a phase of vigorous star formation (SF), fueling the growth of supermassive black holes (SMBHs) and the eventual onset of AGN activity. As the merger progresses, powerful outflows generated by both starburst and AGN winds shape the galactic environment, influencing subsequent SF and black hole accretion with feedback mechanisms (e.g., \citealt{DiMatteo2005, Silk2013}).

Galactic outflows are expected to be more important at high redshift ($z\sim 1 - 2$) during the peak of the cosmic SF and SMBHs accretion (\citealt{Madau2014}), and hence where the feedback mechanisms may be maximized (e.g., \citealt{Brusa2015, Talia2017, Perrotta2019, Kakkad2020, Cresci2023}). Remarkably, recent observations have revealed some similarities between local ULIRGs and distant starbursts at redshift $z\sim2$ (e.g., \citealt{Arribas2012, Hung2014, Bellocchi2022, McKinney2023}) and they  highlight the importance of understanding the outflow phenomena occurring in local ULIRGs. In fact, the proximity of these sources allows for more detailed observations that are characterized by angular resolution and signal-to-noise ratios higher than those focusing on more distant galaxies.




Several studies have demonstrated the presence of multiphase (ionized, neutral, and molecular) gas outflows in ULIRGs (e.g., \citealt{Sturm2011, Bellocchi+13, Veilleux+13, Arribas+14,Feruglio2015, Cazzoli+16, Fluetsch2021, Perna2021, Lai2022, Bohn2024}). 
In this work, we focus on the closest prototypical ULIRG, Arp 220, where evidence of multiphase outflows has already been obtained from several multiwavelength observations (e.g., \citealt{Arribas2001, McDowell2003, Colina2004, Perna2020, Wheeler2020, Lamperti2022, Ueda2022}).

Arp 220 is located at a distance of $\sim 78$ Mpc, and has an IR luminosity of log ($L_{IR}/L_\odot$) = 12.2 (\citealt{Pereira2021}), and a stellar mass of log ($M_{*}$/ $M_{\odot}) \sim 10.8$ (\citealt{U2012}).
It is a late-stage merger containing two compact ($< 150$ pc), highly obscured nuclei (east and west) separated by $\sim$ 370 pc (1\arcsec) that dominate the IR luminosity (e.g., \citealt{Scoville2007}). The two nuclei are sites of intense SF, with SFR $\sim 200 - 250$ \Msunyr (e.g., \citealt{Nardini2010, Varenius2016}) and extremely high SFR surface densities $\Sigma_{SFR} \sim 10^3-10^4$ \Msunyr kpc$^2$, from radio and far-IR (FIR) observations (e.g., \citealt{BarcosMunoz2015, Pereira2021}). There is still no convincing direct evidence for the presence of AGN activity in either of the two nuclei (see \citealt{Perna2024} and references therein).

This paper presents a kinematic examination of the nuclear region of Arp 220 observed with the Integral Field Spectrograph (IFS) unit of the Near Infrared Spectrograph (NIRSpec) instrument on board the James Webb Space Telescope (JWST) (\citealt{Jakobsen2022, Boker2022}) collected as part of the JWST/NIRSpec IFS Guaranteed Time Observations (GTO) survey: Resolved structure and kinematics of the nuclear regions of nearby galaxies (program lead: Torsten B{\"o}ker). Our goal is to investigate the spatially resolved gaseous and stellar kinematics in the innermost 1 kpc of Arp 220, and better understand the feedback processes affecting this ULIRG system.

This paper is organized as follows.
Section \ref{sec:Obs} presents the NIRSpec observations and data reduction. Sections \ref{sec:Analysis_Results} and  \ref{sec:separations} present the spectral fitting analysis and the results through the characterization of the spatially resolved kinematics of the atomic and molecular gas and the identification and study of the different structures found in the central 1.5 $\times$ 1.8 kpc of Arp~220. In Sect. \ref{sec:ISMproperties}, we describe the main properties of the interstellar medium  (ISM). In Sect. \ref{sec:Discussion}, we discuss our results and a comparison with the literature. Finally, Sect. \ref{Conclusions} summarizes our conclusions. Throughout this work, we assume $\Omega_m=0.286$ and $H_0=69.9$ \kms Mpc$^{-1}$ \citep{Bennett2014}. With this cosmology, $1''$ corresponds to 0.368 kpc at the distance of Arp~220.

\begin{figure*}[t!]
    \centering
    \includegraphics[width = 1 \textwidth,trim= 80 0 90 0,clip]{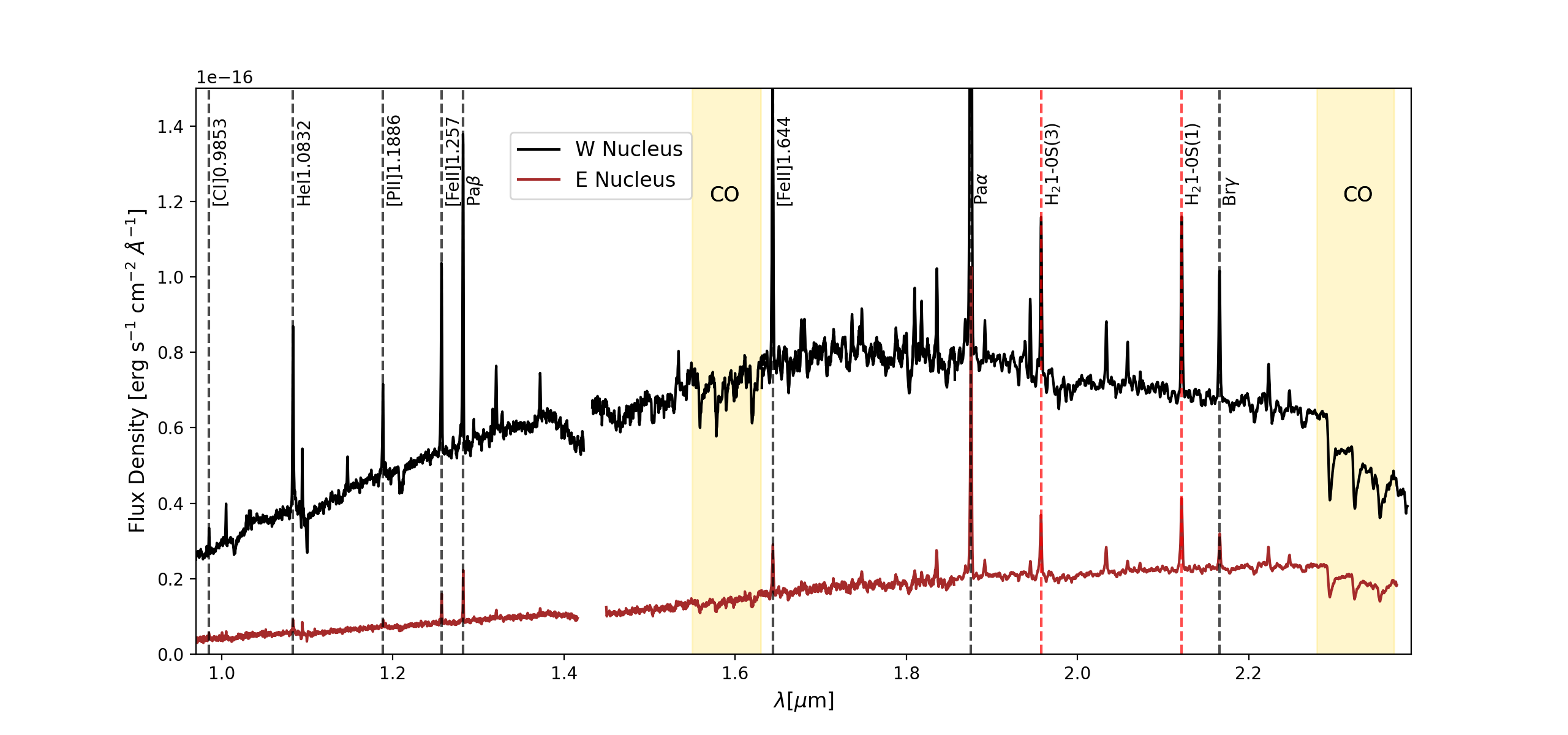}
    \caption{Arp 220  W and E nuclear spectra. The spectra were extracted from a circular aperture of radius 3 spaxels (corresponding to 0.15") and are reported in units of flux density as a function of rest-frame wavelength. Red dashed vertical lines identify H$_2$ lines; black dashed lines identify ionized gas  transitions. Gold-filled areas mark the position of the main CO stellar absorption lines. }
    \label{Integrated Spectra}
\end{figure*}

\section{Observation and data reduction}
\label{sec:Obs}


Arp 220 was observed using the JWST/NIRSpec instrument in IFS mode (\citealt{Boker2022}), as part of the JWST observing program 1267 (PI: D. Dicken), on March 6, 2023. These NIRSpec observations were combined in a single proposal with independent MIRI GTO observations of the same target.

The dataset comprises two distinct pointings, each employing the first four positions of the medium cycling dithering pattern with 15 groups per integration and one integration per exposure, using the NRSIRS2RAPID readout pattern. The first set is centered at the midpoint between the coordinates of the western (W) and eastern (E) nuclei, while the second set is offset by approximately 1\arcsec \ toward the northwest to encompass the \ha\ and \nii\ emission shell (\citealt{Lockhart2015, Perna2020}). Each set has a total integration time of 933 seconds and utilises three high-resolution (R $\sim 2700$) configurations: G140H/F100LP, G235H/F170LP, and G395H/F290LP, covering wavelength ranges of 0.97 -- 1.82 $\mu$m, 1.66 -- 3.05 $\mu$m, and 2.87 -- 5.14 $\mu$m, respectively. However, this paper focuses solely on the first two configurations, as our objective is to study the kinematics of the emitting gas, with its strongest transitions occurring at wavelengths below 2.4 $\mu$m.

The data reduction of Arp 220 observations is described in detail in \citet{Perna2024}. Here we summarize the relevant information. We used the JWST pipeline v1.8.2 with CRDS context 1063. A patch was included to correct some important bugs that affect this specific version of the pipeline; $1/f$ noise and outliers were corrected following the approaches described in \citet{Perna2023a} and \citet{DEugenio2023}, respectively. The combination of the exposure for the eight dither positions was done with a drizzle weighting method that allowed us to sub-sample the detector pixels (\citealt{Law2023}), resulting in cube spaxels with a size of $0.05\arcsec$ (corresponding to $\sim 20$ pc/spaxel) and covering a field of view (FoV) of $5\arcsec \times 4\arcsec$.
The absolute astrometric registration was performed using ALMA high-resolution maps of millimeter continuum emission, as described in \citet{Perna2024}.
The final data cubes showed strong ``wiggles'', sinusoidal modulations caused by the undersampling of the point spread function (PSF). We corrected these artifacts by applying the methodology described in \citet{Perna2023a}.







Figure \ref{RGB_Lines} shows the three synthetic narrow-band images generated from the reduced data cube of Arp~220 obtained for the brightest emission lines in the  wavelength range covered by NIRSpec. 
The positions of the W and E nuclei are marked for visual purposes. This figure illustrates the quality of NIRSpec data and provides a preview of the intricate gas kinematics within the innermost regions of Arp~220. 



\section{Spectral fitting analysis: Stellar and gas components}\label{sec:Analysis_Results}

In this section we describe the spectroscopic analysis carried out to study the kinematics and morphology of the nuclear region of Arp~220. 
We show in Fig. \ref{Integrated Spectra} the integrated spectra of the E and W nuclei up to 2.4 $\mu$m, extracted from a circular aperture of radius 3 spaxels (corresponding to 0.15\arcsec)  centered at the positions defined in \cite{Perna2024}, reported in Table \ref{tab:redshift}. 
The spectra are shown in the rest-frame, according to the systemic redshift of each nucleus: z=0.01840 for the E and z=0.01774 for the W \citep[][see also Sect.~\ref{sec:sistemic_velocity}]{Perna2024}.
The spectra show many hydrogen recombination lines,  including Paschen (Pa) and Brackett (Br) transitions, but also many forbidden iron lines ([FeII]), and roto-vibrational transitions of molecular hydrogen (H$_2$). This allowed us to study the physical and kinematic properties of the gas in different phases. 
The CO absorption bandheads in the spectral ranges 1.5--1.7 $\mu$m and 2.2--2.4 $\mu$m are prominent, and can be used to measure the stellar kinematics.

The analysis is divided into three steps: the first is the modeling and the subtraction of the stellar continuum, the second is a preliminary fit of the emission lines to build a model that can be used to study the integrated properties of each emission line. The third step is the disentangling of the different kinematic components associated with the most prominent emission lines. 
These three steps are described in detail in the following sections.

\subsection{Modeling of the stellar continuum}

To fit the stellar continuum, we used the penalized pixel-fitting routine (pPXF) (\citealt{Cappellarippxf, cappellari2017}) to convolve stellar spectra templates with a Gaussian velocity  distribution.  We did not attempt to extract the Gauss–Hermite terms h3 and h4, since the quality of our data is not sufficient to obtain this information from the fit (see also e.g., \citealt{Cappellari2009, Engel2011, CrespoGomez2021}).
We binned spaxels together using the Voronoi binning scheme of \cite{Cappellari2003}, to achieve a minimum S/N. We established a target minimum S/N $= 30$ per wavelength channel for each bin (following e.g., \citealt{Belfiore2019}).
We used the MARCS stellar population synthesis templates (\citealt{Gustafsson2008}) to model the stellar continuum; these templates cover the wavelength range 0.96 -- 20 $\mu$m with a constant spectral resolution $\Delta\lambda/\lambda$ = 20000. We also adopted a third-order multiplicative and additive polynomials to better reproduce the spectral continuum shape. 
To improve the quality of the fit, we included in the total model the most prominent emission lines (see Table \ref{tab:lines}) parameterized with two Gaussian components. 

\begin{figure}[t!]
    \centering
    \includegraphics[width = 0.5 \textwidth]{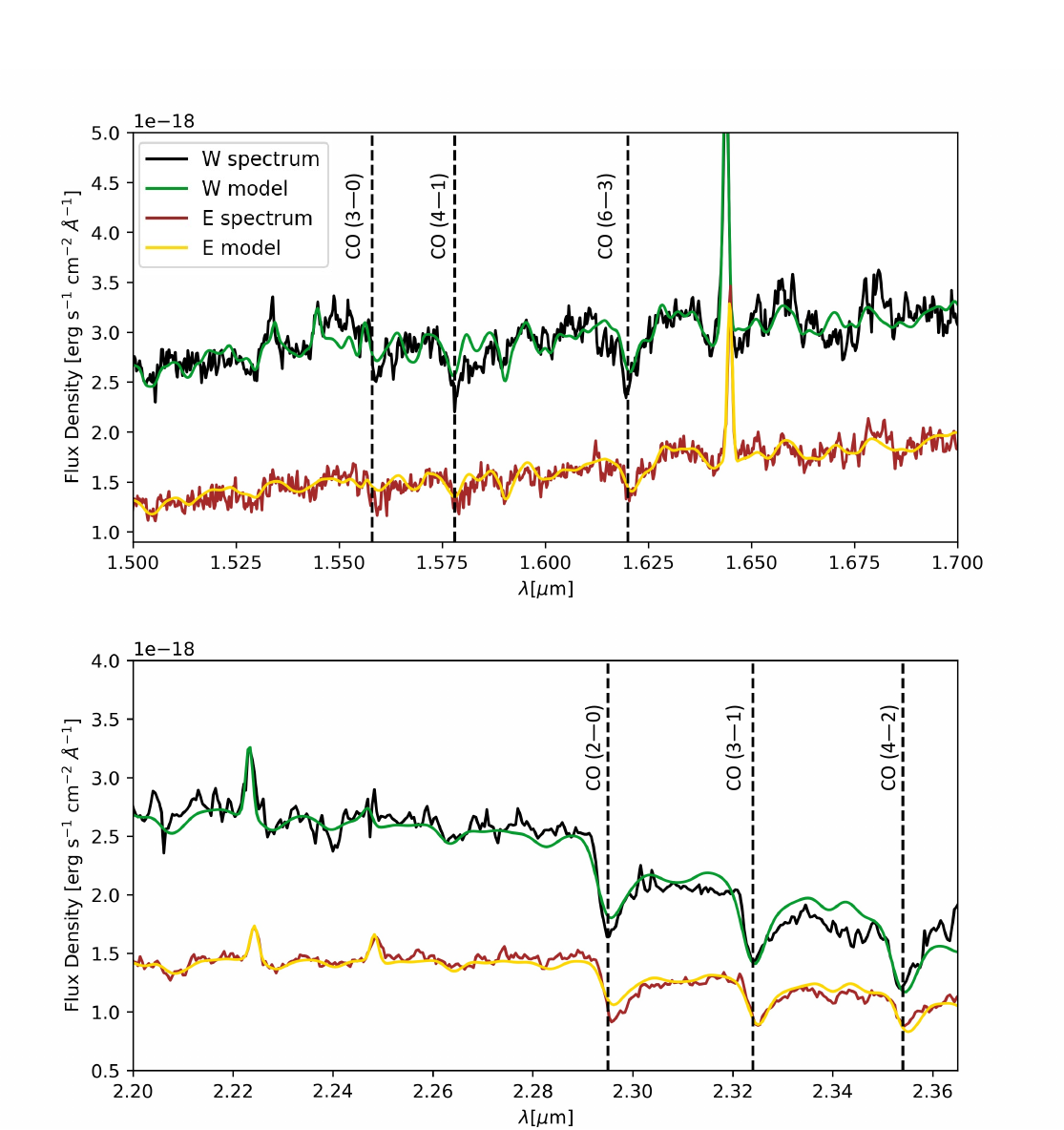}
    \caption{ Arp 220  W and E nuclear spectra with pPXF best-fit models. The CO absorption features are in the 1.5--1.7~$\mu$m band (upper panel) and 2.2--2.36~$\mu$m (bottom panel). Black and green curves respectively refer to the data and the model of the W nucleus spectrum. Brown and gold curves respectively refer to the data and the model of the E nucleus spectrum. Black dashed vertical lines indicate the position of the main CO absorption lines.  }
    \label{ppxf fit nuclei}
\end{figure}

We conducted separate fits for the cubes G140H and G235H, with the latter restricted to wavelengths below 2.4 $\mu$m, to avoid spaxels affected by the NIRSpec detector gap. Wavelengths falling within the gap in G140H were appropriately masked during the fit procedure.
Fig. \ref{ppxf fit nuclei} shows the fit with pPXF of the spectra extracted from the W and E nuclei, highlighting the most important features in absorption (black dashed lines) to constrain stellar kinematics. We note that the pPXF best fits does not perfectly reproduce the observed spectra. However, this does not affect
our goal: we do not require a realistic model of the stellar populations in Arp 220, we only need an accurate stellar continuum model and a measurement of the stellar kinematics.

After constructing the total model, we subtracted the model of the continuum from the cube spaxel by spaxel, rescaling the modeled continuum emission obtained in each Voronoi bin to the median flux of the observed continuum in each spaxel. In this way, we obtained a cube containing only the contribution of the emission lines.

\subsection{Multi-Gaussian emission line fit }
\begin{table}[b]
\centering
\caption{Fit emission lines. Vacuum wavelengths are taken from \cite{Perna2024}.}%
\begin{tabular}{cc|cc}
\hline
Atomic Gas & $\lambda_{vac}$ & Hot Molecular Gas  & $\lambda_{vac}$\\
  &  ($\mu m$) &  & ($\mu m$) \\
\hline
[CI] & 0.9824 & H$_2 1-0~S(7)$ & 1.7481\\
& 0.9853 & H$_2 1-0~S(6)$ & 1.7880\\
Pa$\gamma$& 1.0941 & H$_2 1-0~S(5)$ & 1.8359\\
Pa$\beta$ & 1.2821 & H$_2 1-0~S(4)$ & 1.8920\\
Pa$\alpha$ & 1.8756 & H$_2 1-0~S(3)$ & 1.9577\\
Br$\gamma$ & 2.1661 & H$_2 1-0~S(2)$ &  2.0339\\
HeI & 1.0832 & H$_2 1-0~S(1)$ & 2.1220\\
 & 1.8690 & H$_2 1-0~S(7)$ &2.2235\\ 
 & 1.8702 &  &\\

$[$PII$]$ & 1.1471 &  &\\
& 1.1886 &  &\\

$[$FeII$]$ & 1.2570 \\
& 1.3725 \\
& 1.3209 & &\\
& 1.2787 & &\\
& 1.6440 & &\\

[OI] & 1.3168 &  &\\

\end{tabular} 
\label{tab:lines}
\end{table}

After the subtraction of the continuum, we smoothed each wavelength map of the cube (i.e., image generated with a spectral pixel wide synthetic filter) with a Gaussian having a dispersion of 1 spaxel (i.e., 0.05\arcsec), to enhance the S/N.
We then performed separate spectral fits for the atomic and molecular gas line groups listed in Table \ref{tab:lines} since each gas phase may be dominated by different kinematic components, as illustrated in Fig. \ref{RGB_Lines} (see also e.g., \citealt{May2017, Perna2024}). All lines within each group were simultaneously fitted using one, two, or three Gaussian components.
We tied the velocity $v$ and the velocity dispersion $\sigma$ of each Gaussian component among all lines of the group, while the fluxes are left free to vary. Then, we chose the number of Gaussian components for each spaxel by performing a Kolmogorov-Smirnov (KS) test on the distribution of the residuals with $N$ and $N+1$ components to define the minimum number of components required to provide an acceptable fit in each spaxel (\citealt{Marasco2020}).
Consequently, the number of Gaussian components is not based on a physical model but it is constrained on just a statistical analysis; a detailed, physical decomposition is presented in Sect. \ref{sec:separations}.

\begin{figure*}[t!]
    \centering
    \includegraphics[width = 1 \textwidth]{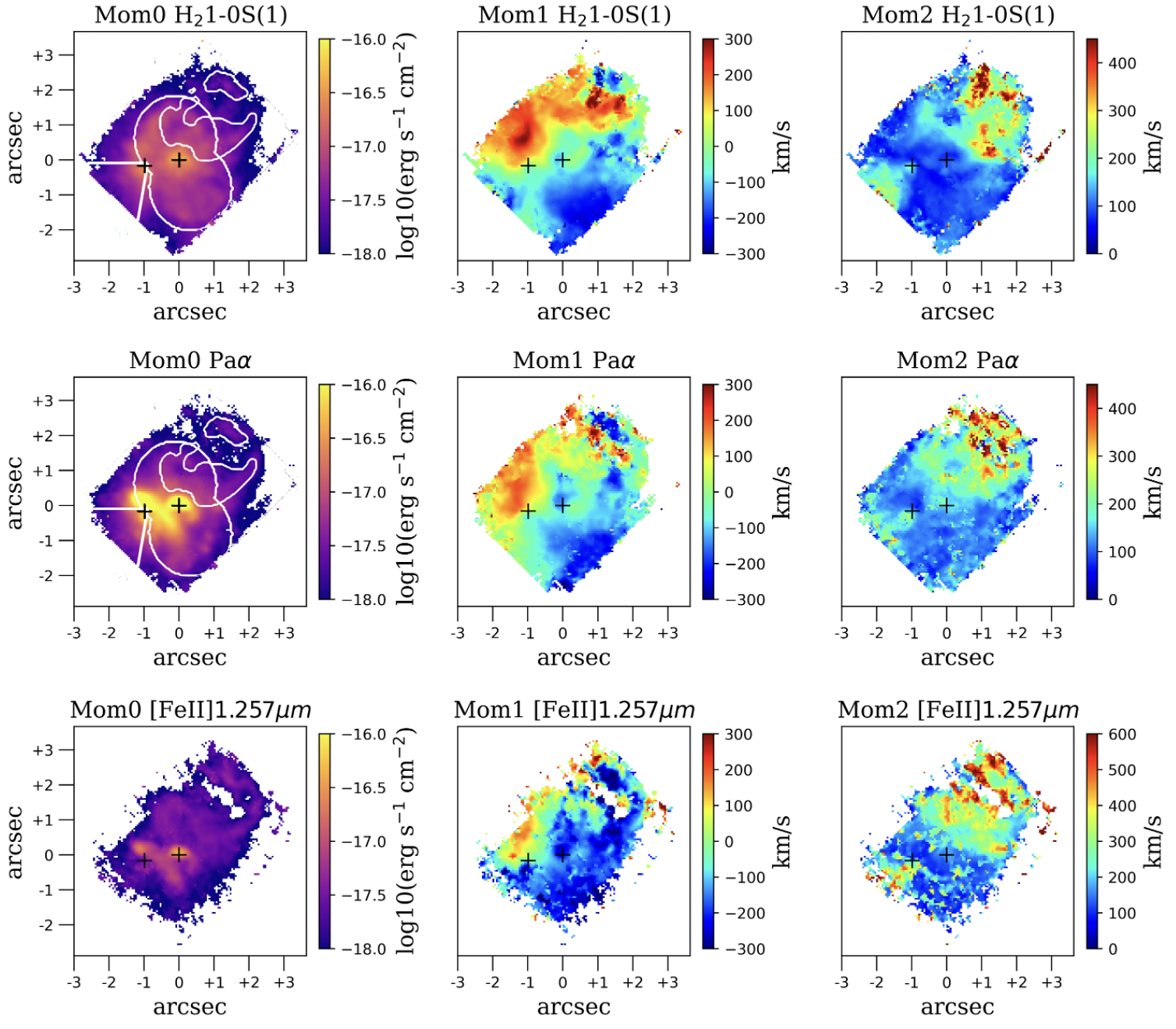}
    \caption{ Moment maps of hot molecular gas (H$_{2}$~1-0S(1), top) and ionized gas (center: Pa${\alpha}$, bottom: \feii~1.257$\mu$m). The moment-1 was computed with respect to z$_m$ = 0.0181. An S/N cut of three on the flux of each emission line was applied to the maps. The black "+" symbols mark the position of the E and W nuclei. White contours represent the edges of the masks described in Sec. \ref{sec:separations}. } 
    \label{Moments}
\end{figure*}

Figure \ref{Moments} shows the three moment maps calculated on the whole modeled line profile (consisting of either one, two, or three Gaussians) of the brightest emission lines: H$_2~1-0~S(1)~2.122~\mu$m, Pa$\alpha$, and \feii~1.257. For each transition, we masked spaxels with S/N$<$3.
These maps allowed us to identify the different kinematic and morphological components of the system, traced by ionized and molecular gas, and described in Sect.~\ref{sec:separations}. 

\subsection{Systemic velocity}
\label{sec:sistemic_velocity}
Following \citet{Perna2024}, we determined the redshift of the two nuclei from the centroid of the atomic hydrogen lines in the nuclear spectra. The systemic velocities correspond to $z_E^{gas}$ = 0.01840 $\pm$ 0.00001  and $z_W^{gas}$ = 0.01774 $\pm$ 0.00001. We also computed the redshift from the stellar continuum model, finding $z_E^*$ = 0.0183 $\pm$ 0.0002 and $z_W^*$ = 0.0179 $\pm$ 0.0003, slightly different from the redshift found for the gas, but still consistent within the errors (see Table \ref{tab:redshift}).

\begin{table}[t]
\centering
\caption{Coordinates and systemic redshift of the two nuclei of Arp 220 from \citet{Perna2024}.}%
\begin{threeparttable}
\begin{tabular}{lcc}
\hline
  & W nucleus & E nucleus \\
\hline
RA (J2000) & 15:34:57.224  & 15:34:57.294  \\
DEC (J2000) & +23:30:11.515 & +23:30:11.353\\
redshift (gas) & 0.01774 $\pm$ 0.00001 & 0.01840 $\pm$ 0.00001 \\
redshift (stars) & $0.0179\pm 0.0003$ & $0.0183\pm 0.0002$ \\
\hline
\end{tabular} 
  \end{threeparttable}
\label{tab:redshift}
\end{table}

In this work, the large-scale ($> 100$ pc) stellar and gas kinematics maps are derived by using as a reference the redshift reported in \cite{Perna2020}, which corresponds to the median value between $z_E^*$ and $z_W^*$, $z_m = 0.0181$ (e.g., Fig. \ref{Moments}). 
This redshift allows us to obtain a symmetric stellar velocity gradient, associated with the  kiloparsec-scale disk also identified with optical and cold molecular line tracers (see \citealt{Perna2020} and references therein).
In contrast, the small-scale motions of the ionized and hot molecular gas around the E or W nucleus are measured relative to the zero velocity inferred from $z_E^{gas}$ and $z_W^{gas}$, respectively.


\begin{figure*}[t!]
    \centering
    \includegraphics[width = 1 \textwidth]{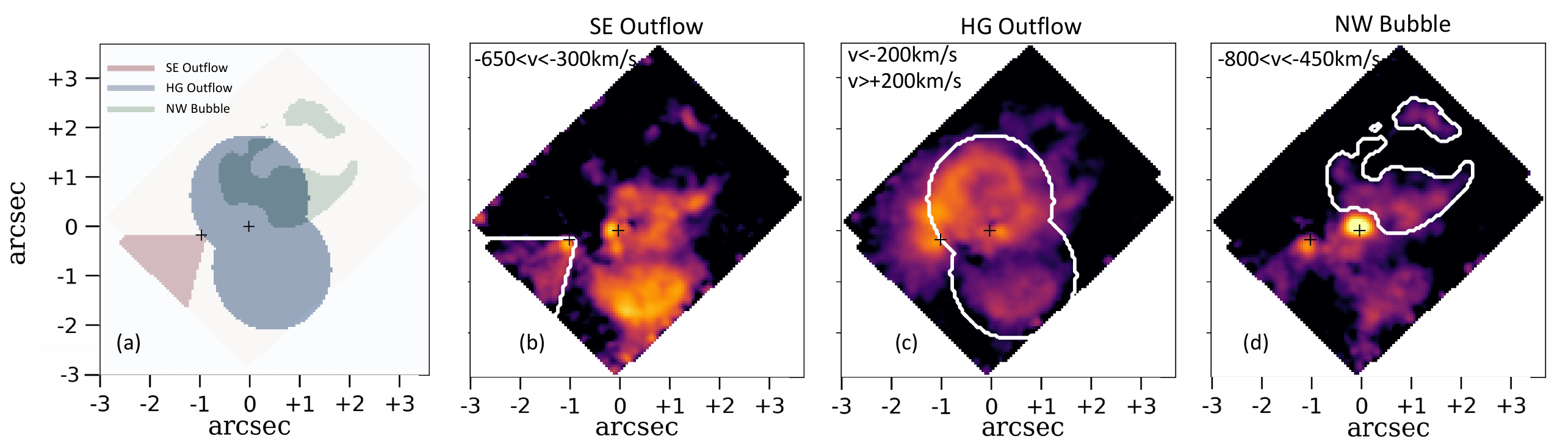}
    \caption{ Arp 220 outflow regions. (a): Composite view of the isolated outflow regions: southeast outflow (red), hour-glass outflow (blue), northwest outflow (green).
    (b): Channel map between $-650$ and $-300$~\kms in the H$_{2}$ transition  with respect to the systemic redshift of the E nucleus. The white curve indicates the edges of the SEO mask.
    (c): Sum of channel maps $<-200$ and $>200$ \kms in the H$_{2}$ transition  with respect to the systemic redshift of the W nucleus. The white curve indicates the edges of the HG region.
    (d):  Sum of  H$_{2}$ and Pa$\alpha$  channel maps between $-800$ and $-450$ \kms  with respect to the systemic redshift of the E nucleus. The white curve indicates the edges of the NWB region.}
    \label{Mask_Features}
\end{figure*}

\section{Spectral fitting analysis: Separating rotations and outflows components}\label{sec:separations}

The morphology and kinematics of the nuclear region of Arp~220 are quite complex. Figs. \ref{RGB_Lines} and \ref{Moments} reveal a superposition of 
circum-nuclear and large-scale disks as well as multiple outflows that have already been observed in optical and submillimeter tracers (e.g., \citealt{Scoville2017, Perna2020, Lamperti2022, Ueda2022}). 
Emission lines often display multipeak and asymmetric spectral profiles, requiring the use of multiple Gaussian components for modeling. 
Conventional methods usually employed to separate a disk from an outflow structure across the entire galaxy (e.g., based on velocity or velocity dispersion criteria; \citealt{Tozzi2021,Venturi2021}), may not be straightforward for Arp 220. Hence, in this work we opted for an alternative approach, focusing on individual regions where the separation of distinct kinematic components is facilitated by dedicated criteria and by the high spatial resolution of NIRSpec data. In particular, we first selected and modeled the most easily distinguishable outflow components, subtracted them, and then gradually move on to the more complex ones.


Our method consists of three steps. 
The first is the identification of a region with specific morphology and/or kinematics, and hence the construction of a spatial mask to isolate it. 
We then fitted all spaxels within the selected region with N Gaussians, where N represents the number of the kinematically distinct components we wanted to characterize. As the fit of a complex line profile is subject to degeneracy, the value N is quite critical and depends on the detectability of the morphological and kinematic features we aim to describe. 
We considered in our model the most intense transitions of the atomic (Pa$\alpha$) and molecular gas (H$_{2}~ 1-0~S(1)~ 2.122\ \mu$m, hereafter H$_2$ unless otherwise specified) in the G235H cube. We also included in the fit the HeI~1.869~$\mu$m and He~1.870~$\mu$m lines since they can contaminate the profile of Pa$\alpha$ (see, e.g., Fig. B.3 in \citealt{Perna2024}).
We fitted together the transitions of atomic and molecular gas assuming for each line the same $v$ and $\sigma$ and varying the fluxes. As we  show in the next sections, this is a reasonable assumption for this system.
In the second step, we defined a criterion to isolate an outflow among the components identified in the mask. 
Then, we constructed a model cube containing only the isolated component, used for further analysis presented in the following sections. 

This three-step process is iterative, implying that each time we aimed to characterize a specific feature, we initially created a mask and then fitted a new model on the residual cube built by subtracting all previously detected and isolated components. The criteria used to define all masks and outflow components are introduced in the next sections (Sect.~\ref{sec_Ose}--\ref{WOutflow}).

Fig. \ref{Mask_Features}a illustrates the masks for all the identified outflow regions. They encompass a conical outflow launched by the E nucleus (i.e., the southeast outflow, SEO hereinafter), an hourglass-shaped outflow originating from the W nucleus (HGO), and a northwestern bubble (NWB). There is another feature of particular interest that affects just a few spaxels and is therefore not reported in the figure: a compact outflow observed in the W nucleus mainly in the ionized phase (WNO).
In the next subsections, we provide detailed analyses of each identified region. 


\begin{figure*}[t!]
    \centering
    \includegraphics[width = 1\textwidth]{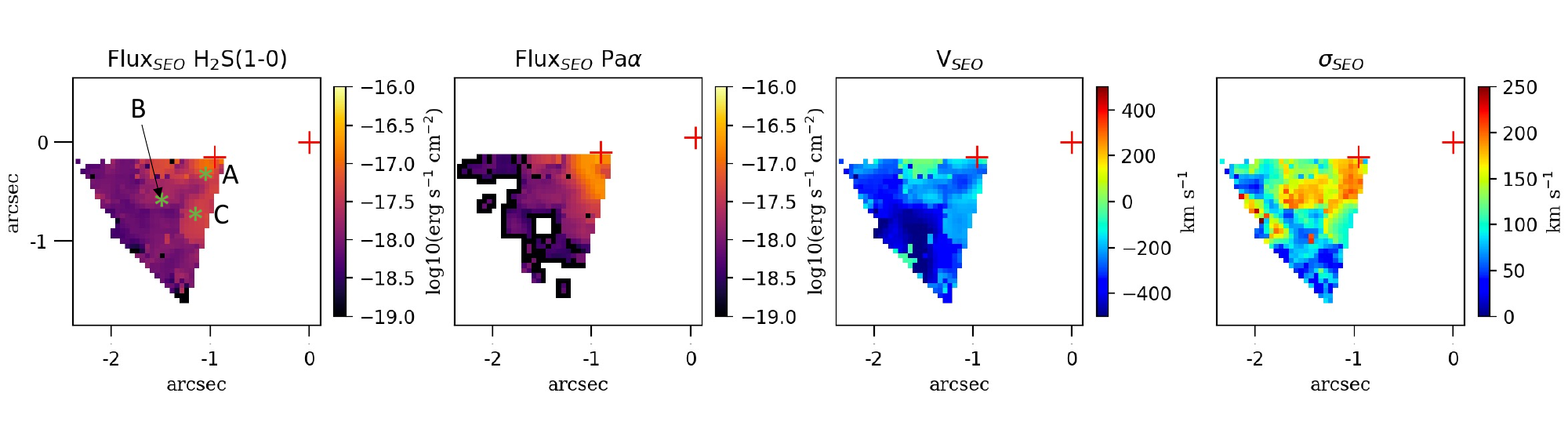}
    \caption{SE conical outflow flux and velocity maps. From left to right: flux of H$_{2}$ and Pa$\alpha$ lines, velocity, and velocity of 
    the Gaussian component attributed to the SE outflow. The velocities are relative to the systemic velocity of the E nucleus. Red crosses indicate the positions of the two nuclei. Green asterisks in the first panel indicate the spaxels presented in Fig. \ref{SE_Outflow Spectra}.  }
    \label{SE_Outflow}
\end{figure*}

\begin{figure*}[t!]
    \centering
    \includegraphics[width = 0.95\textwidth]{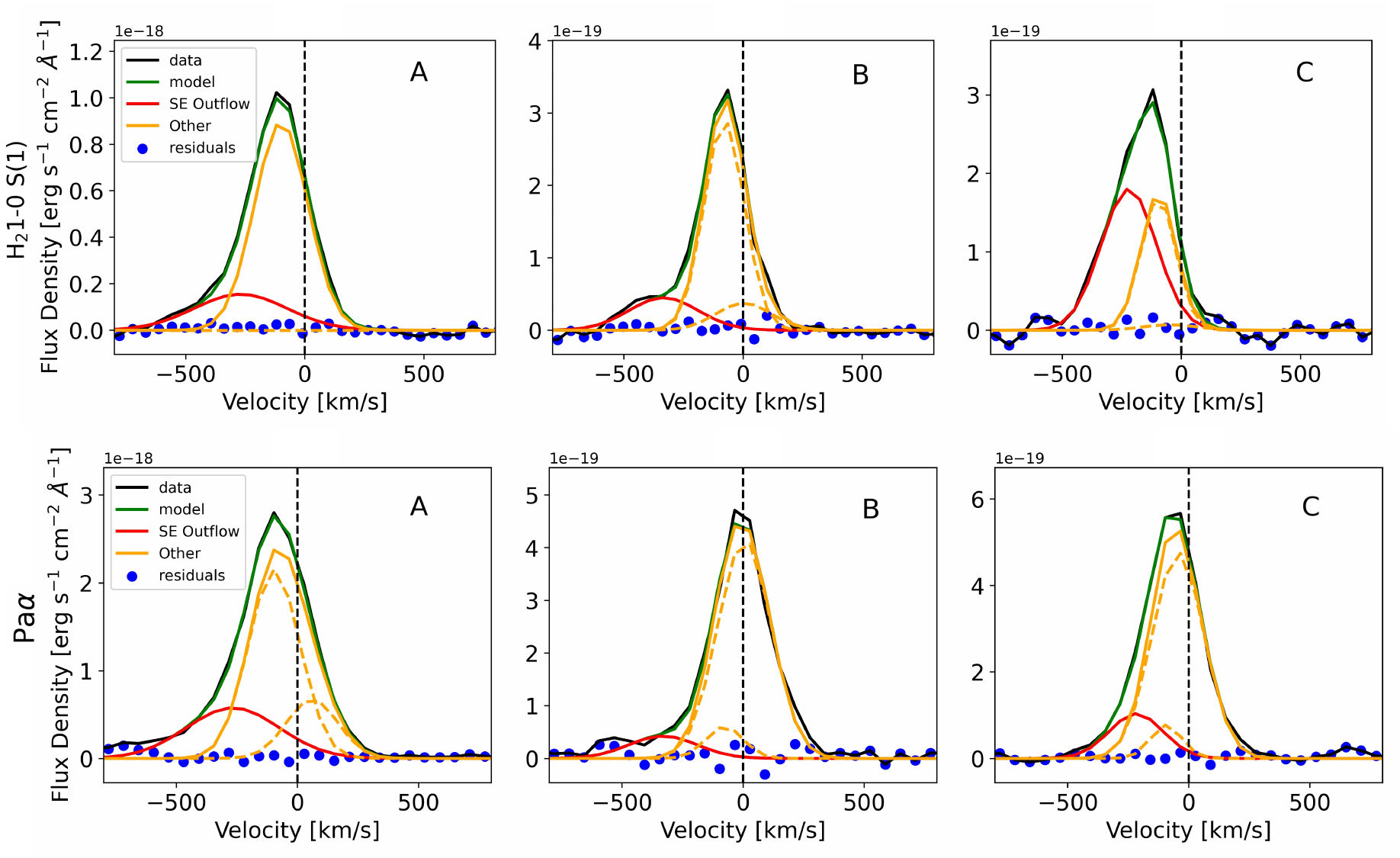}
    \caption{Spectra extracted from the SE outflow cone. The solid black curves represent the line profiles of H$_{2}$ (top) and Pa$\alpha$ (bottom) extracted from the spaxel close to the E nucleus (A), in the central part of the cone (B), and at the edge of the collimated outflow (C), as labeled in Fig.~\ref{SE_Outflow}. 
    The multi-Gaussian fits are depicted by dashed orange curves for the disk components and solid red curves for the outflow component. The total model for the disk is shown as a solid orange curve, and the combined model for both disk and outflow is displayed in green. 
    The blue dots indicate the residuals. The velocities are computed with respect to the systemic velocity of the E nucleus. }
    \label{SE_Outflow Spectra}
\end{figure*}

\begin{figure*}[t!]
    \centering
    \includegraphics[width = 1 \textwidth]{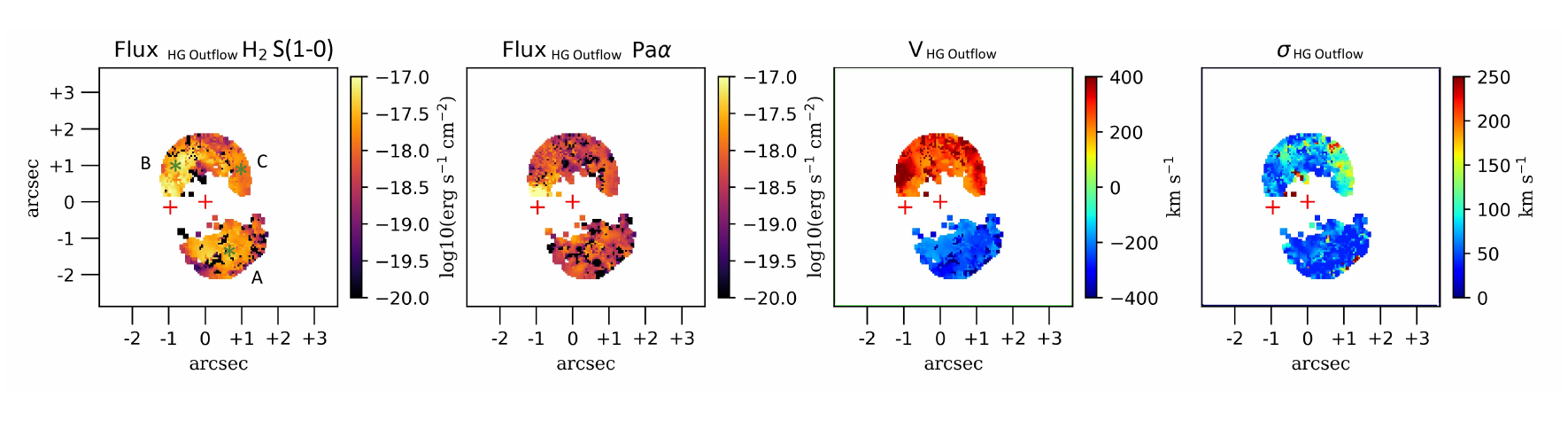}
    \caption{HG outflow flux and velocity maps. From left to right: flux of H$_{2}$ and Pa$\alpha$ lines, velocity, and velocity dispersion of the Gaussian component attributed to the HG outflow. The velocities are relative to the systemic velocity of the W nucleus. Red crosses indicate the positions of the two nuclei. Green asterisks in the first panel indicate the spaxels presented in Fig. \ref{fig:Hourglass_fit}.}
    \label{Hourglass}
\end{figure*}
\begin{figure*}[h!]
    \centering
    \includegraphics[width = 0.95 \textwidth]{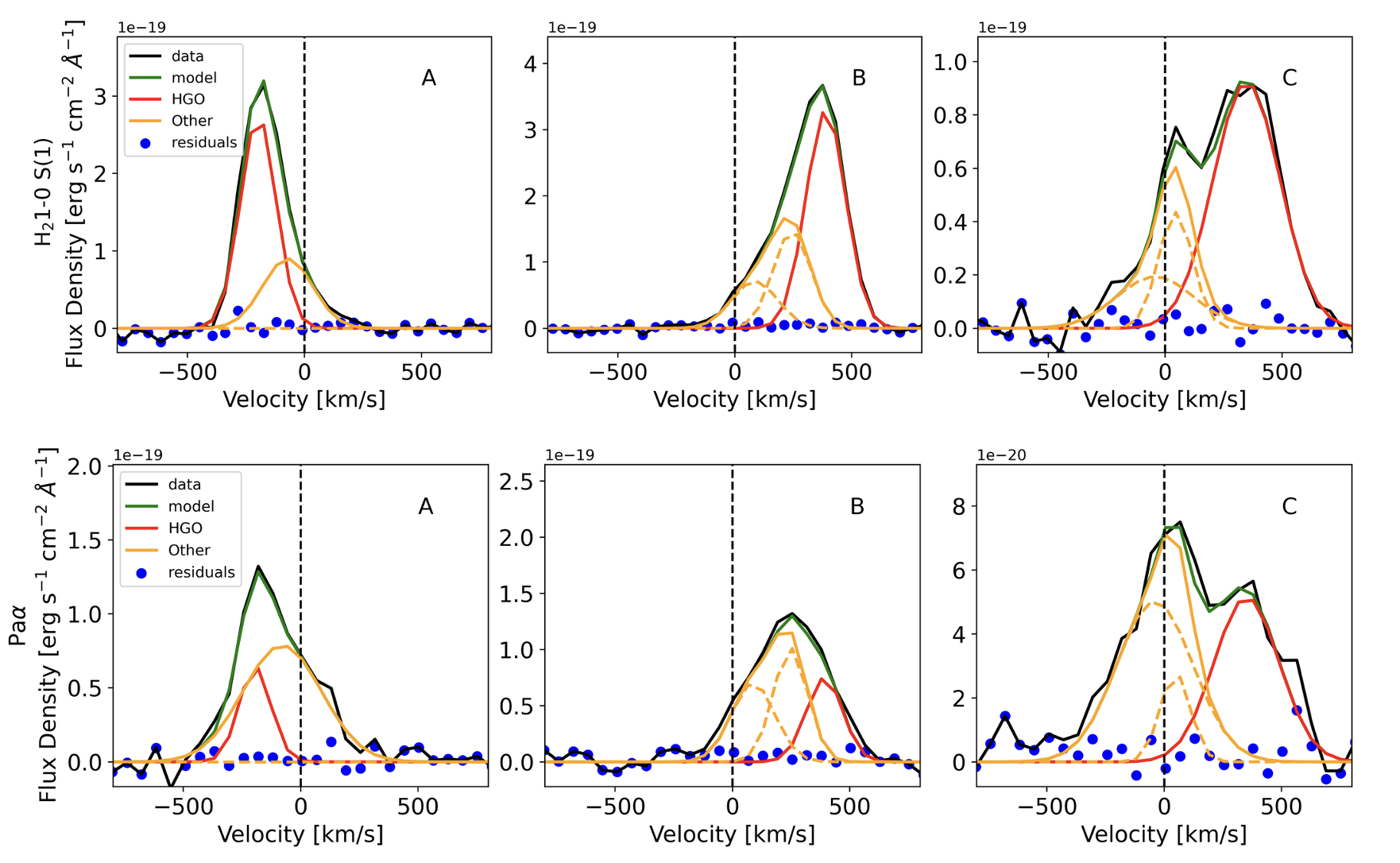}
    \caption{Spectra extracted from the HG outflow region. The solid black curves represent the line profiles of H$_{2}$  (top) and Pa$\alpha$ (bottom) extracted from the spaxels marked in Fig. \ref{Hourglass}. 
    The red curve refers to the HG outflow. Orange curves identify the additional components used for the fit (orange dashed curves). The combined model is displayed in green. 
     The velocities are computed with respect to the systemic velocity of the W nucleus.}
    \label{fig:Hourglass_fit}
\end{figure*}

\subsection{Southeast conical outflow from E nucleus}\label{sec_Ose}


Figure \ref{Mask_Features}b shows H$_{2}$ emission in the velocity range $[-650, -300]$~\kms with respect to the systemic velocity of the E nucleus. In this figure, a high-velocity conical region situated southeast of the E nucleus is clearly visible (demarcated by white lines).  This region has also elevated velocity dispersion ($>150$~\kms), as shown in the moment-2 maps of both ionized and hot molecular gas in Fig. \ref{Moments}, hence indicating the presence of an outflowing component.

We selected the region with enhanced flux in the $[-650, -300]$~\kms map (red area in Fig.~\ref{Mask_Features}a) and, within this region, fitted both the ionized gas and the molecular gas with three Gaussian components to model: 1) the high-velocity gas, 2) the rotating disk at a large scale that covers the entire FoV, and 3) the emission from the small circum-nuclear disk around the E nucleus. 
Among these three Gaussian components, we assigned the most blueshifted to the  outflow. 
The flux, the velocity and the velocity dispersion of the SEO in  H$_{2}$ and Pa$\alpha$ are reported in Fig. \ref{SE_Outflow}. 
Since we associated only one Gaussian component to the outflow, the velocity and the velocity dispersion is the same for both ionized and hot molecular gas. 
The velocity varies between $-100$ \kms, observed close to the nucleus and at the edges of the cone, and $- 400$~\kms at larger distances ($\sim 400$~pc). High velocity dispersions (up to 200 \kms) are observed in the center of the conical outflow and close to the nucleus, while at the edges of the cone $\sigma$ is about two times lower.

Figure \ref{SE_Outflow Spectra} shows the spectra of the hot molecular (top) and ionized gas (bottom) extracted from three spaxels within the conical outflow; specifically near the E nucleus, at the center, and at the edge of the cone. In all spaxels, we observe a prominent wing extending toward negative velocities; this wing is reproduced by a Gaussian component with $v < -250$~\kms, which we associated with the SEO (red curves in the figure). We also note that the SEO emission is more prominent and contributes more to the total line profile in the hot molecular phase than in the ionized one, especially in the off-nuclear regions. Figure \ref{SE_Outflow Spectra} also shows that our multi-Gaussian approach is  capable of simultaneously reproducing the profiles of ionized and hot molecular gas, hence supporting the validity of our assumption about the presence of a mixed, multiphase ISM in each kinematic component. 


As explained in Sect. \ref{sec:separations}, our approach consists of an iterative process: the best-fit model component attributed to the SEO is subtracted spaxel by spaxel from the original data cube before performing new fits to characterize the remaining kinematic structures observed in the NIRSpec data. This method helps in reducing the fit degeneracy.


\subsection{Hourglass outflow from the W nucleus}\label{sec:HGoutflow}


As shown in Fig. \ref{RGB_Lines}, both the ionized and molecular gas exhibit a shell-shaped morphology at high redshifted ($v > 200$~\kms) velocities 
in the region north of the W nucleus. Additionally, the hot molecular gas shows a blueshifted shell-like structure in the region southwest of the same nucleus. The blue- and redshifted gas forms an `hourglass' (HG)-like shape, 
compatible with an outflow launched from the W nucleus. This structure is even clearer in Fig. \ref{Mask_Features}c, showing the sum of H$_{2}$ spectral channel maps at $v< -200$~\kms and $v> 200$~\kms (with respect to the systemic velocity of the W nucleus), used to  
define our mask (white contours in the figure). 

Within the HG region, we fitted simultaneously the hot molecular and the ionized gas using three Gaussian components, taking into account the large- and small-scale gas rotations as well as the outflow.
To assign a Gaussian component to the HG outflow (HGO), we discarded the components with absolute velocities $< 180$~\kms. Among the remaining ones, we attributed to the outflow the component with highest absolute velocity. The threshold (180 \kms) represents a compromise: with a smaller velocity threshold, we would have included in the outflow also part of the rotational disk at larger scales; with a higher value, we would have missed part of the contribution of the outflow.

Figure \ref{Hourglass} shows the flux, velocity and velocity dispersion of the gas component attributed to the outflow.
The flux is dominated by the hot molecular gas over the ionized one. The velocity field is rather symmetrical with respect to the W nucleus and varies between $-300$ and 400~\kms. 
The highest observed velocity is located to the east in the redshifted shell, corresponding to an increase in molecular gas emission line flux. It is unclear whether this high-velocity gas is still part of the HGO, 
or whether it may represent a redshifted counterpart of the conical SE outflow associated with the E nucleus (Sect. \ref{sec_Ose}). 

Figure \ref{fig:Hourglass_fit} shows the spectra extracted from three spaxels within the HG region (A, B, C in Fig. \ref{Hourglass}), to illustrate our fitting procedure and the separation of the kinematic components. 
The red curve refers to the component attributed to the outflow, while the orange curves refer to the other components used for the fit. We note that the line profiles in this region are very broad and multiple components are needed to model them. 
Nevertheless, despite differences in the line profiles of the hot molecular and ionized gas at specific positions, the multi-Gaussian fits successfully reproduce these profiles, confirming the reliability of our fit decomposition.



\begin{figure*}[t!]
    \centering
    \includegraphics[width = 1 \textwidth]{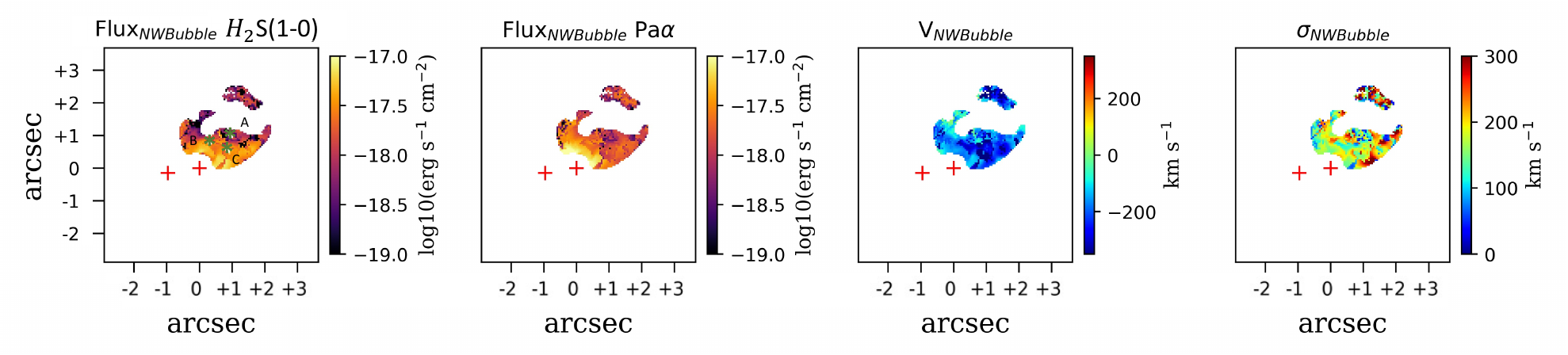}
    \caption{NW bubble flux and velocity maps. From left to right: flux of H$_{2}$ and Pa$\alpha$ lines, velocity, and velocity dispersion of the Gaussian component attributed to the outflow. The velocity is referred to the E nucleus. Red crosses indicate the positions of the two nuclei. Green asterisks in the first panel indicate the spaxels presented in Fig. \ref{spectra:Northern_Bubble_Fit}. }
    \label{Northern_Bubble}
\end{figure*}

\begin{figure*}[h!]
    \centering
    \includegraphics[width = 0.95 \textwidth]{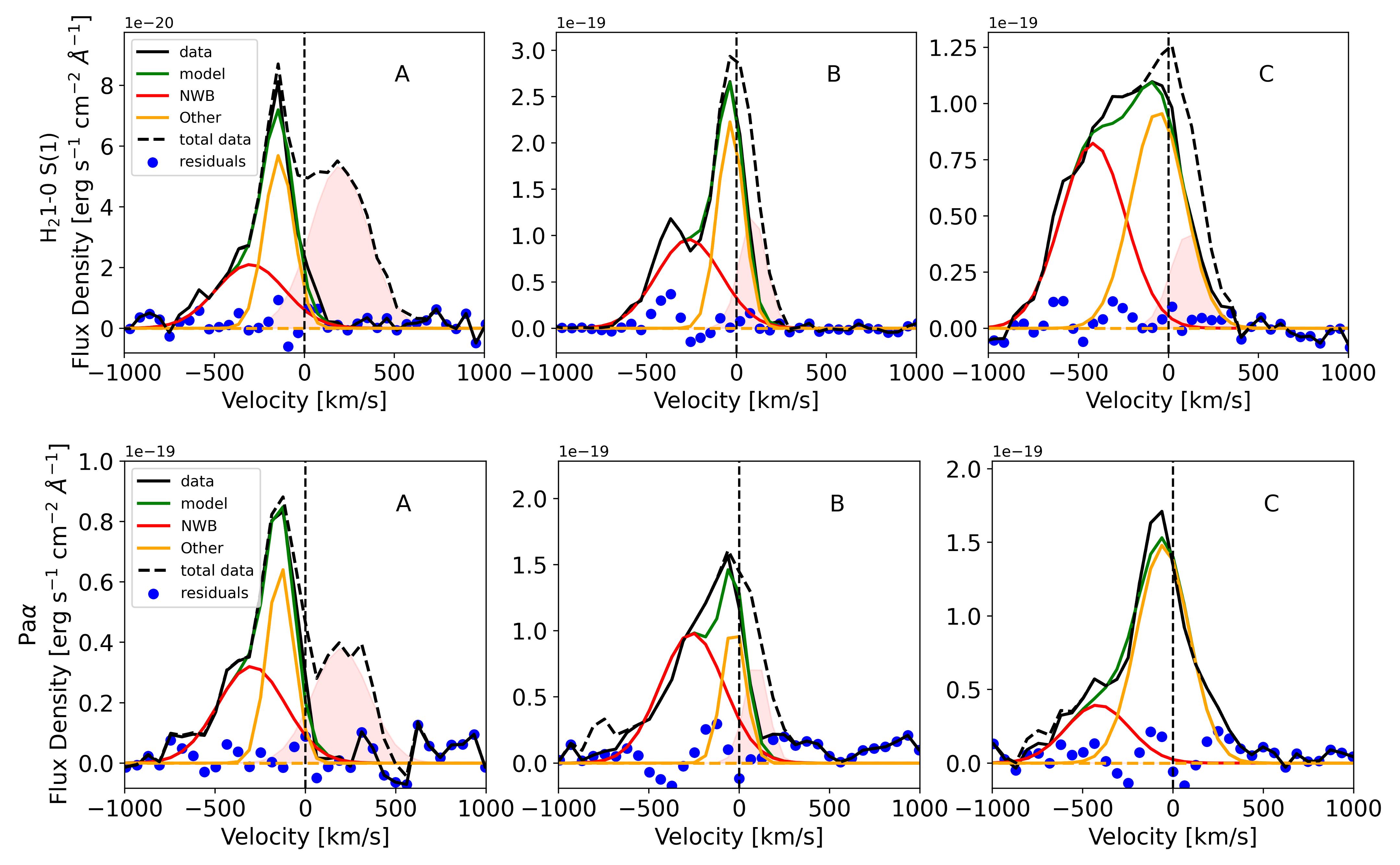}
    \caption{Spectra extracted from the NW bubble region. Line profiles of H$_{2}$ (top)  and Pa$\alpha$ (bottom) are shown at the position of the three spaxels labeled in Fig. \ref{Northern_Bubble}.  The black dashed curves represent the spectra from the original cube, highlighting the contribution attributed to the HGO (red-shaded area) that we have already isolated and subtracted; the black solid lines represent the spectra fitted to identify the NW bubble. The red Gaussian profiles identify the isolated NW bubble, while orange profiles identify the rotating disk component. The combined model is displayed in green. 
    The velocities are computed with respect to the systemic velocity of the E nucleus.}
    \label{spectra:Northern_Bubble_Fit}
\end{figure*}

\begin{figure*}[t!]
    \centering
    \includegraphics[width = 0.97 \textwidth]{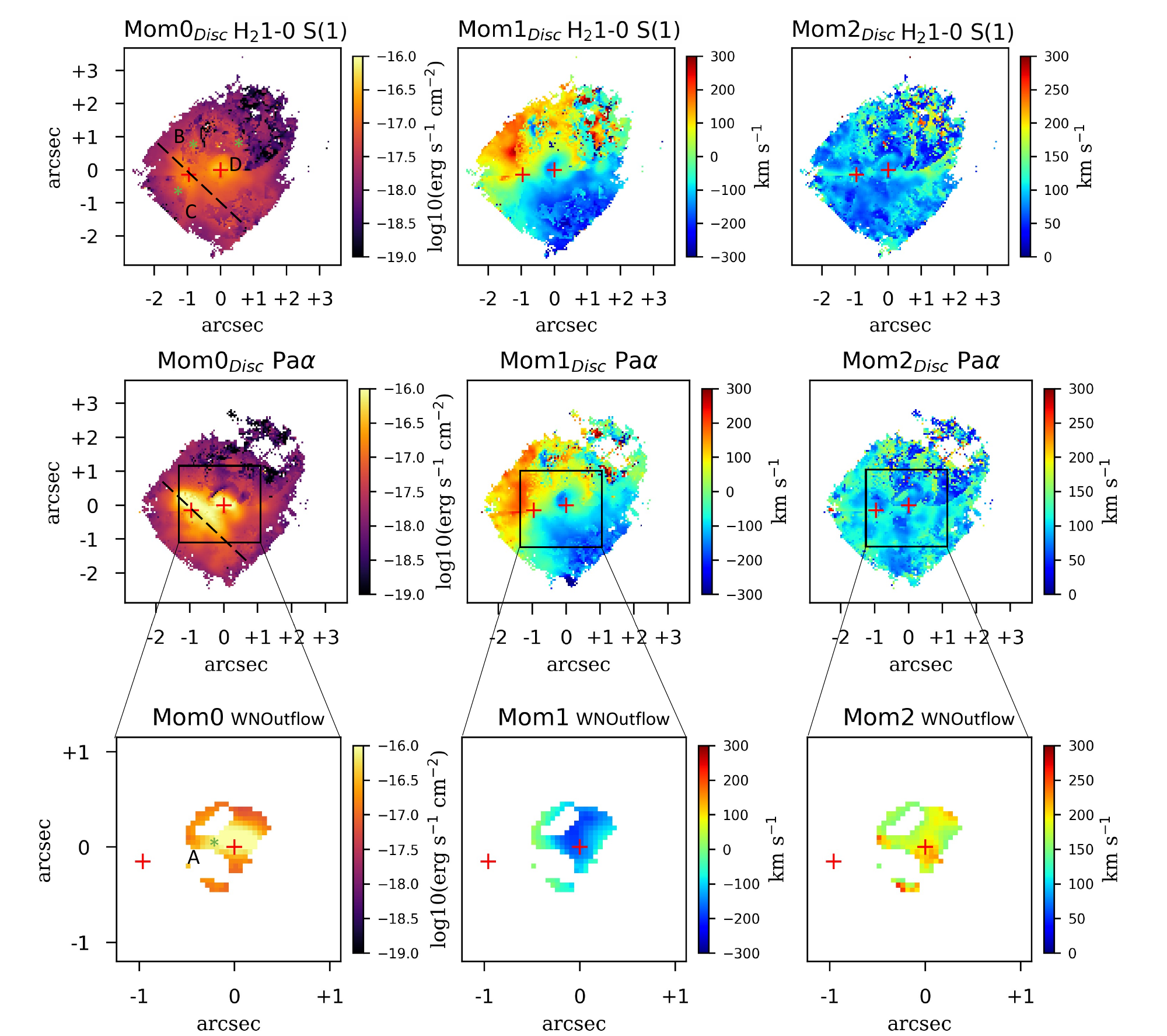}
    \caption{Moment maps of the large-scale rotating disk and the W nuclear outflow. Top and middle panels: moment-0 (left), moment-1 (center), and moment-2 (right) maps of H$_{2}$1-0S(1) (top) and Pa$\alpha$ (middle) of the Gaussian components attributed to the perturbed disk. The black-dashed line in the velocity maps show the large-scale disc mean kinematic major-axis (PA = 45°). 
    Bottom panels: Pa$\alpha$ moments maps of the Gaussian  components attributed to the W nuclear outflow.
    Velocities are referred to z$_{m}$ in the first two rows. Velocity in the zoom-in inset is referred to the systemic velocity of the W nucleus.}
    \label{rotational disc}
\end{figure*}

\begin{figure}[h!]
    \centering
    \includegraphics[width = 0.5 \textwidth]{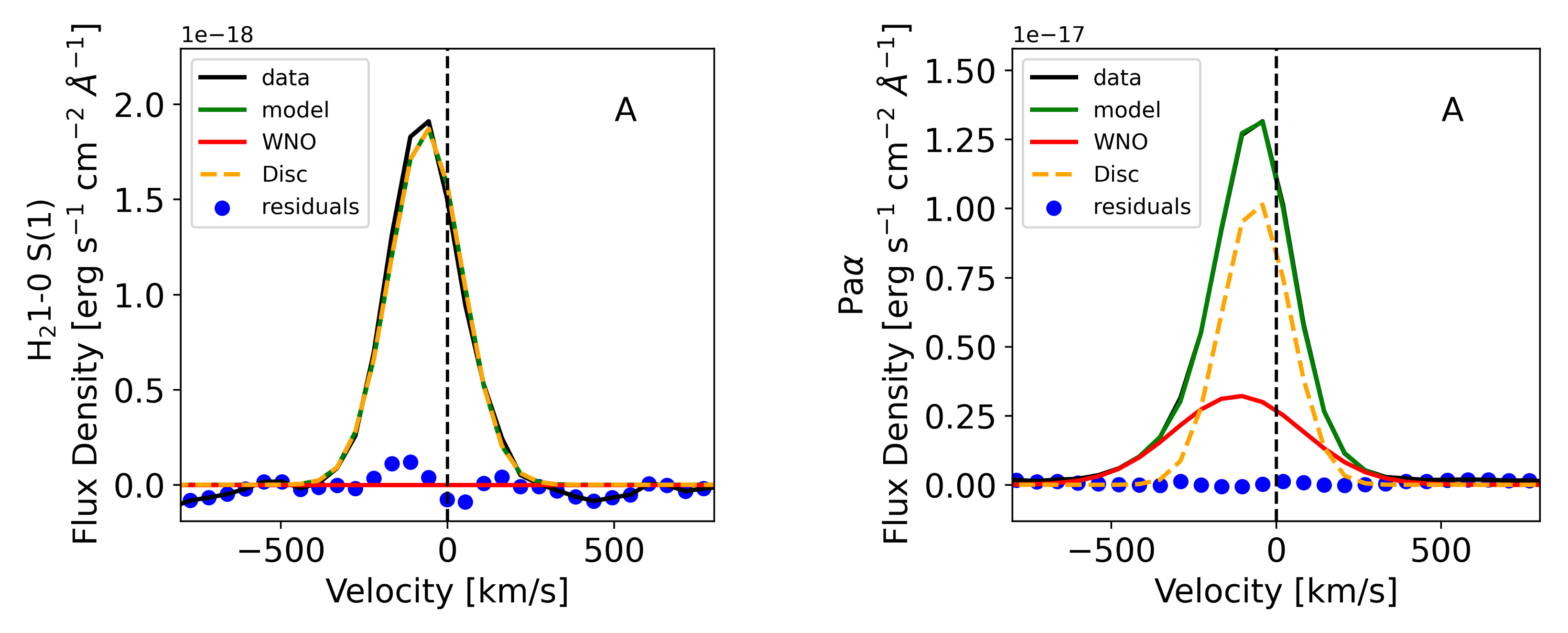}
    \caption{Spectrum extracted from the WNO. Line profiles of H$_{2}$ (left) and Pa$\alpha$ (right) extracted from a spaxel labeled in the bottom panel of Fig. \ref{rotational disc}). 
    The two component fits are represented by the dashed orange (disc component) and red (outflow component) curves; the combined model is displayed in green. The velocities are computed with respect to the systemic velocity of the W nucleus.}
    \label{fig:spectraWout}
\end{figure}

\subsection{Northwest bubble}
\label{sec:NWBubble}

We identified a large scale bubble in the north west of the NIRSpec FoV, mostly visible in the \feii transitions 
(Fig. \ref{RGB_Lines}). Figure \ref{Mask_Features}d displays the sum of spectral channel maps at  $v<-450$~\kms (with respect to the systemic velocity of E nucleus) of Pa$\alpha$ and H$_{2}$.  We created the mask for the NW bubble (NWB) by selecting those spaxels with enhanced flux in these velocity channels, excluding the region around the W nucleus (white curve in the figure).
This nuclear region was already included in Sect.\ref{sec:HGoutflow}, but  since its outflow kinematics are slightly different from those of the HG outflow, this region is discussed in Sect. \ref{WOutflow}.  

The emission that contributes to the line profiles in this mask comes from the bubble and the large-scale rotating disk. Consequently, we fitted the emission lines with two Gaussian components. No specific boundaries on velocity and velocity dispersion were used during the fit. 
We assigned the most blueshifted component to the outflow. This Gaussian component also exhibits a higher velocity dispersion with respect to the one associated with the disk.

Figure \ref{Northern_Bubble} shows the flux, velocity and velocity dispersion of the gas component attributed to the outflow. In this region, the fluxes of Pa$\alpha$ and H$_{2}$ are very similar, except in the area closest to the W nucleus, where the ionized phase dominates. The velocity ranges between $-100$ and about $-450$~\kms, with velocities dispersion $> 150$~\kms.

Figure \ref{spectra:Northern_Bubble_Fit} shows the fit of Pa$\alpha$ and H$_{2}$ in three spaxels extracted from the bubble. 
The black dashed curve represents the spectra from the original cube, highlighting the component (red-shaded area) attributed to the HGO that we have already isolated and subtracted. The red curves represent the NWB component, while the orange curves correspond to the other components used in the fit.
In all panels, the nature of the Gaussian components is clear: the outflow component is blueshifted and broader than the disk component.


\subsection{Compact outflow from the W nucleus} 
\label{WOutflow}
There is another kinematic feature in the W nuclear region that has not been isolated yet. It is a nuclear outflow (WNO) that is visible in the ionized gas phase (see also \citealt{Perna2024}). 
However, it affects just a few spaxels and cannot be isolated in the moment maps (Fig. \ref{Moments}) nor in the velocity channel maps (e.g., Fig. \ref{RGB_Lines}). 
Therefore, once all the features described before (i.e., the SEO, the HGO, and the NWB)  have been subtracted, we fitted the residual cube with two components to model the WNO and the perturbed rotating gas (i.e., large-scale rotation over the entire FoV, or small-scale rotating nuclear discs). 

We attributed to the WNO the components with $\sigma$ > 150 \kms  within a radius of 0.4" (150 pc) around the nucleus. In the more external regions, 
we associated both the components to the perturbed rotating gas. 
Figure \ref{rotational disc} shows the three moments of the components attributed to the perturbed disk in H$_2$ and Pa$\alpha$ (top and central panels) and of the WNO  (bottom panels).
The WNO kinematics are characterized by blueshifted velocities up to $-200$~\kms and a velocity dispersion of $\sim 150-200$~\kms. 

Figure~\ref{fig:spectraWout} shows the spectrum extracted from a random spaxel (marked in the bottom panel of Fig. \ref{rotational disc}) to illustrate the fit decomposition for the ionized and hot molecular gas phases.  The differences in the line profile of the two gas phases are evident: 
the broad, blueshifted wing that is observed in Pa$\alpha$ is not observed in the molecular gas. 
\begin{figure*}[t!]
    \centering
    \includegraphics[width = 0.95 \textwidth]{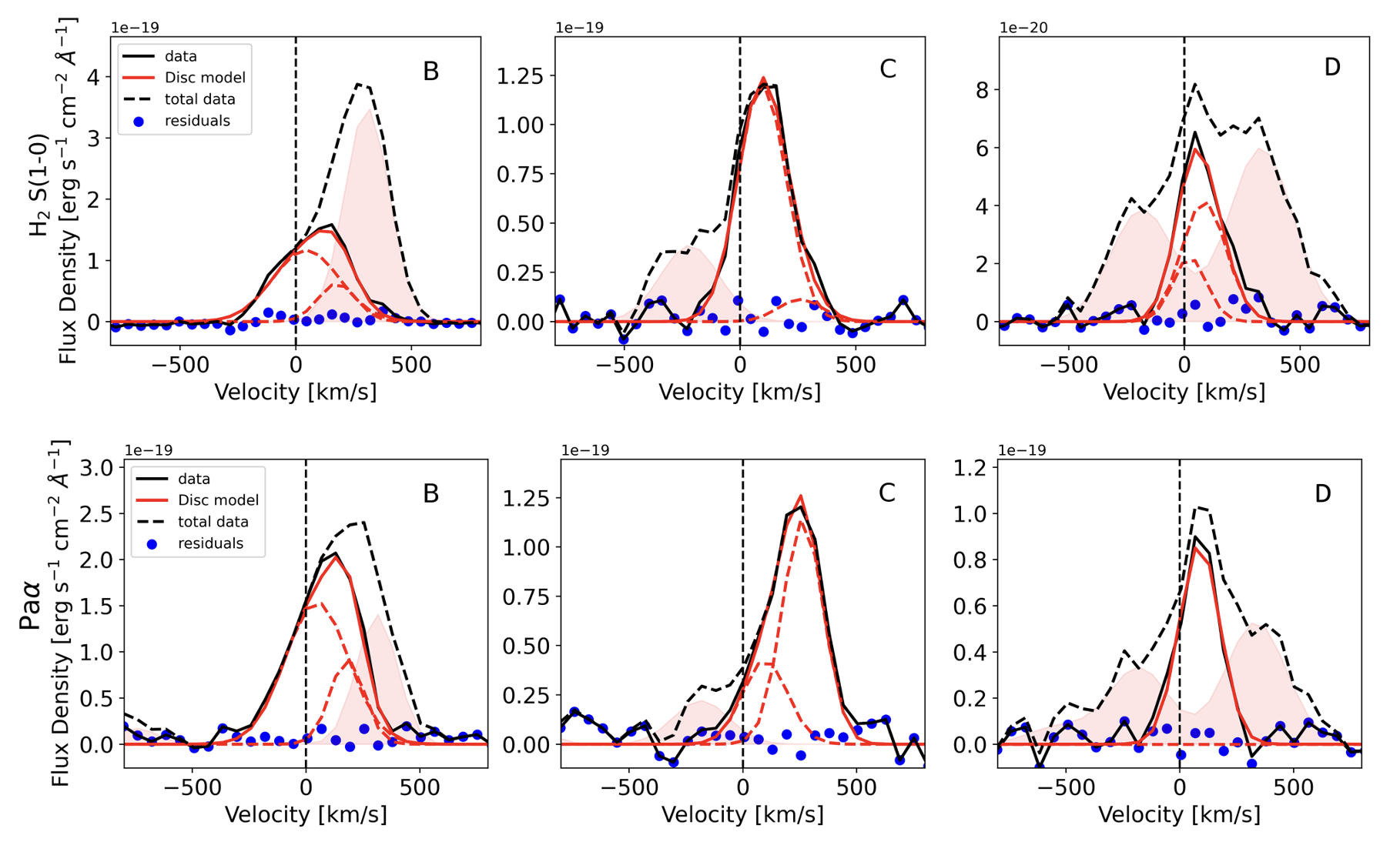}
    \caption{Spectra extracted from the kiloparsec-scale disk. Line profiles of H$_{2}$ (top)  and Pa$\alpha$ (bottom) at the position of the three spaxels labeled in Fig. \ref{rotational disc}.  The black dashed curves represent the spectra from the original cube, highlighting the contribution attributed to the HGO and SEO (red-shaded area) that we have already isolated and subtracted; the black solid lines represent the spectra fitted to identify the large-scale disk. 
    The red dashed curves show  the components associated with the disk model. The total disk model is shown with a solid red curve. Velocities are computed with respect to z$_{m}$. }
    \label{fig:Disc_spectra}
\end{figure*}

\subsection{Perturbed rotating disk}
\label{Perturbed_Rotating_Disc}
The moment maps of hot molecular and ionized gas 
shown in Fig. \ref{Moments} already suggested a large-scale rotation pattern, although strongly affected by several outflow features. These additional kinematic components were isolated and subtracted in Sects. \ref{sec_Ose}, \ref{sec:HGoutflow}, \ref{sec:NWBubble} and \ref{WOutflow}. 
The fitting procedure used to separate the compact nuclear outflow (WNO) component from the rotating disk is described in the previous section (Sect.~\ref{WOutflow}). After the subtraction of  the components associated with the WNO,  we assigned to the perturbed rotating gas all the components having $\sigma < 200$~\kms. In fact, 
the components with higher velocity dispersion might represent residuals of outflowing gas that we did not model in the previous steps. 
Figure~\ref{rotational disc} (top and central panels) shows the moment maps of the isolated perturbed rotating gas, with contributions from both circum-nuclear discs in the vicinity of the E and W nuclei and the large-scale disk. The Moment-1 velocities are defined with respect to $z_m = 0.0181$, which is more appropriate for the kiloparsec-scale disk (see Sect. \ref{sec:sistemic_velocity}). 

We observe a symmetric velocity gradient with respect to the E nucleus, with velocities from $-250$~\kms (northeast) to $+250$~\kms (southwest). The Moment-1 map in the northwest part of the FoV is instead disturbed due to the non-perfect subtraction of HGO and NWB contributions. Moreover, significant deviations from a large-scale rotational patter are also observed in the vicinity of the W nucleus: this is due to the presence of a small-scale disk surrounding the W nucleus (see also e.g., \citealt{Scoville2017}). 
We calculated the position angle  (PA)\footnote{The position angle is measured on the receding half of the galaxy, taken anticlockwise from the north direction on the sky (east of north).} of the major kinematic axis of the large-scale disk by fitting the Moment-1 map of Pa$\alpha$ and for H$_{2}$ with the FIT KINEMATIC PA code  \citep{Cappellari2007, Krajnovic2011}, obtaining 51$\pm$7° and 39$\pm$7°, respectively. 
In Sect.~\ref{sec:stellar_kinematics}, we  use the average of these two values to construct the position-velocity (PV) diagram. 


The Moment-2 of the perturbed rotating disk reaches values up to 150 \kms, significantly lower than in the moment 2 map obtained before the subtraction of the outflowing features (Fig. \ref{Moments}).
Moreover, the kinematics of Pa$\alpha$ are more perturbed compared to that of the hot molecular gas, as also seen in \cite{Armus2023}.

Figure~\ref{fig:Disc_spectra} shows three single-spaxel spectra extracted from the HG region (B), the SEO cone (C), and an overlapping region of the HG and NW bubble (D) (See Fig.~\ref{rotational disc}).
After the subtraction of the outflow components, the line profiles are now rather symmetrical; the second component used in the fit procedure (red dashed curves) either is much fainter than the first component, or shows similar velocity shifts and line widths. Therefore, the attribution of both components to the disk is well justified. 

\subsection{Stellar kinematics}
\label{sec:stellar_kinematics}
\begin{figure*}[t!]
    \centering
    \includegraphics[width = 0.95 \textwidth]{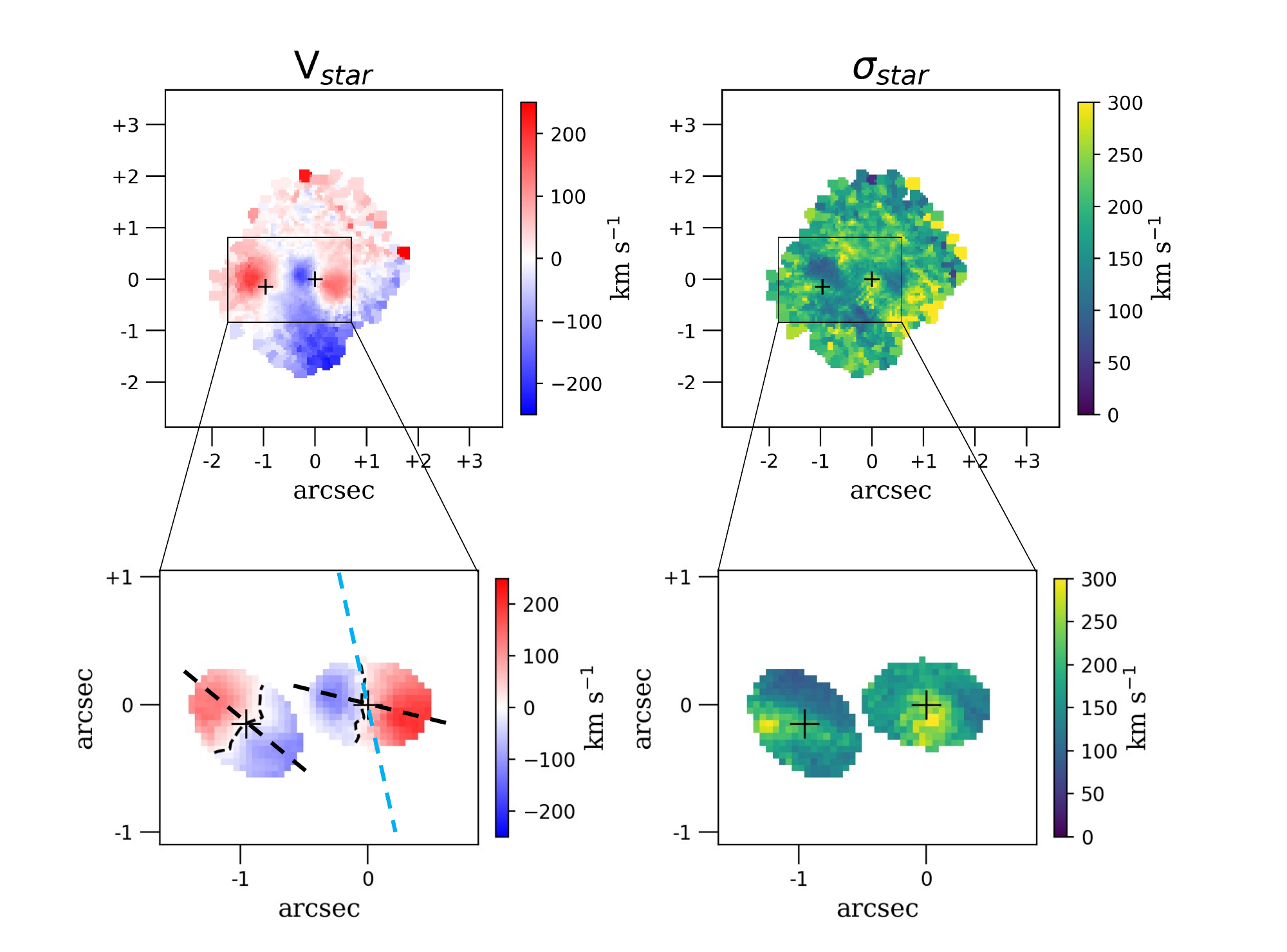}
    \caption{Stellar kinematics maps. Top panels: Stellar velocity (left, in the  frame z$_m$) and velocity dispersion (right). Bottom panels: Stellar velocity (left) and velocity dispersion (right) around the two nuclei shifted with respect to the rest-frame velocity of each nucleus. Black dashed lines refer to the major axis of the two nuclei (W: PA = 257°, E: PA = 45°). Light blue dashed line refers to the axis of HG outflow (PA = 13°). }
    \label{Stellar_Kinematics}
\end{figure*}

Figure \ref{Stellar_Kinematics} presents the stellar kinematics maps obtained with pPXF. In particular, they are created from the analysis of the G235H cube, which is less affected by dust attenuation than the G140H cube and shows more prominent CO stellar absorption features in the wavelength range 2.2--2.4~$\mu$m  (Fig. \ref{Integrated Spectra}). 
In Fig.~\ref{1cube_stellar} we show that the maps derived from the G235H and G140H gratings are consistent.
As shown in the inset panel of Fig. \ref{Stellar_Kinematics}, the stellar velocity field of Arp 220 presents two nuclear discs rotating in opposite directions separated by 1'', with a velocity offset of $\sim$ 120~\kms (see also \citealt{Wheeler2020,Sakamoto1999}). 
The stellar velocity dispersion is higher in the nuclear positions, and then decreases throughout a region crossing the two nuclei, in correspondence with the bright elongated region observed in the Pa$\alpha$ Moment-0 map (Fig. \ref{rotational disc}).
Moreover, in the E nucleus, the velocity dispersion increases toward the east, while in the W nucleus it increases in the south direction. 
In the outermost regions north of the two nuclei, the velocity dispersion reaches about 200 \kms, consistent with the values found in \cite{Perna2020}. 

In the next two sub-sections we compare the stellar and gas kinematics making use of PV diagrams.


\subsubsection{Stellar and gas kinematics around the W nucleus}
\begin{figure*}[h!]
    \centering
    \includegraphics[width = 0.9 \textwidth]{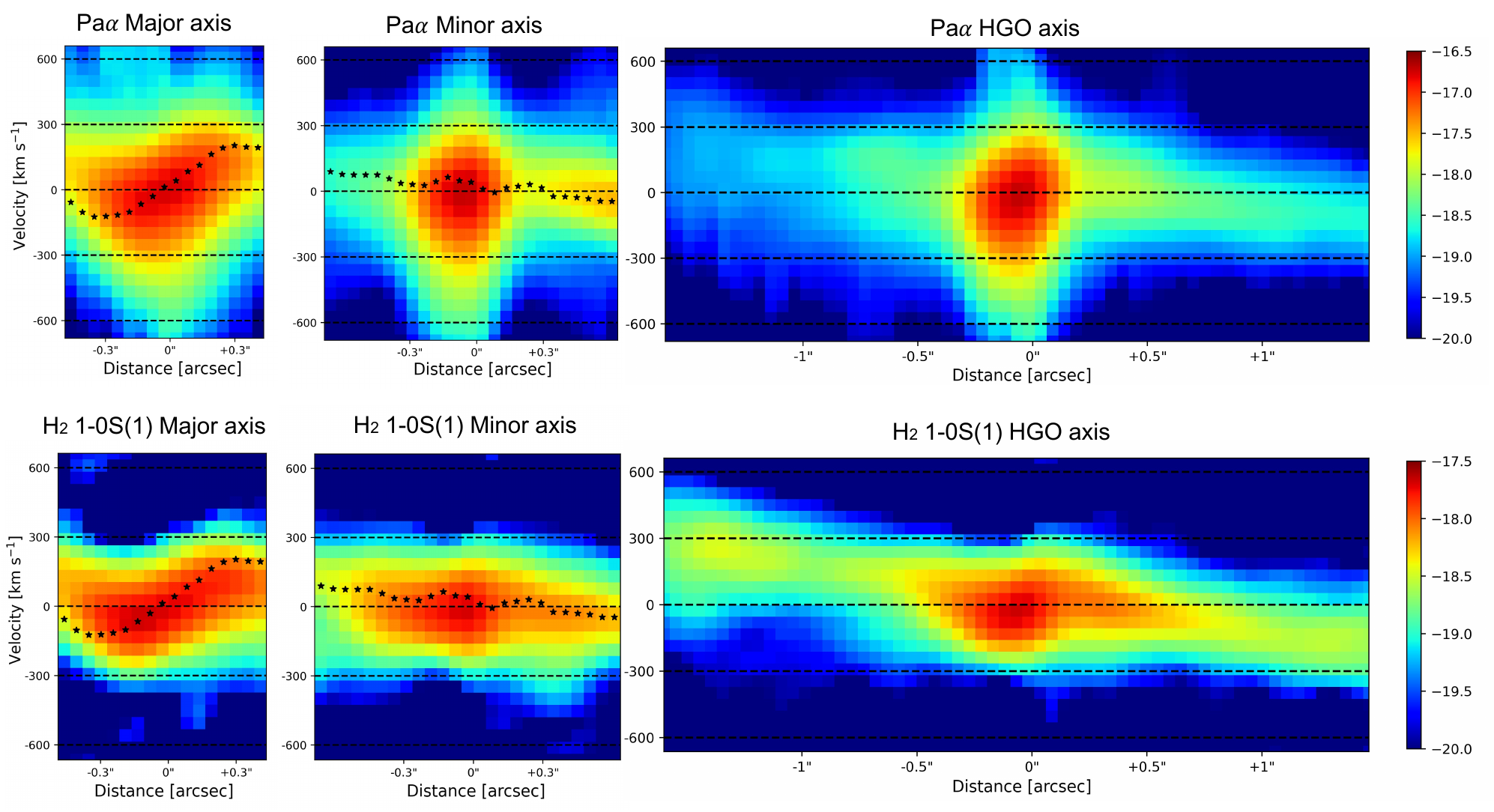}
    \caption{Position-velocity diagram for ionized gas (top) and hot molecular gas (bottom) of W nucleus with positions varying along the disk major axis (PA = 257°, left, with positive distances toward west), the disk minor axis (PA = 347°, center, with positive distances toward north); and HG outflow axis (PA = 13°, right, with positive distances toward southwest). Black stars represent the stellar velocity. The zero-velocity refers to z$^{*}_{W}$.  }
    \label{pv_diagram_west_nucleus}
\end{figure*}

In this section we compare the gas and stellar kinematics around the W nucleus. 
We constructed a PV diagram of both the ionized gas and the hot molecular gas (Fig. \ref{pv_diagram_west_nucleus}), centered on the W nucleus with a PA $\sim$ 257$\pm$6° directed along the major axis of rotation of the circumnuclear stellar disk calculated with the FIT KINEMATIC PA code (dashed black line in Fig.~\ref{Stellar_Kinematics}). We set the zero velocity at the redshift of the W nucleus ($z_{W}^{*}$). The ionized gas (upper panel) reaches  blueshifted velocities higher than those of the hot molecular gas (bottom panel). The stellar component (black star symbols in the plots) follows the kinematics of ionized gas, while it appears to rotate faster than the hot molecular gas. 

In the PV diagram along the circumnuclear disk's minor axis  (PA $\sim$ 347$\pm$6°), the ionized outflow (WNO), discussed in Sect.~\ref{WOutflow}, is evident. It extends about 0.3" from the W nucleus and exhibits velocities reaching up to 600~\kms both in the blue- and redshifted sides. However, this feature is not detected in the hot molecular gas, probably due to the dissociation of the H$_{2}$ molecules or, more in general, to different physical conditions (e.g., in terms of magnetic field and electron density in the outflowing gas; see, e.g., \citealt{Kristensen2023}). 

We also constructed a PV diagram along the axis of the HG outflow (PA = 13°, left panel of Fig. \ref{pv_diagram_west_nucleus}), which is not exactly perpendicular to the axis of rotation of the circumnuclear disk. Ionized gas dominates the emission within 0.5" radius of the W nucleus, and decreases considerably at larger distances; instead, the H$_{2}$ emission is more diffuse.
The high blue- and redshifted velocities spans a range of $-200$~\kms$<v<300$~\kms at a distance of about 1.5" (600 pc), delimiting the HG outflow. 
To summarize, around the W nucleus  we found 1) a relatively good match between stellar and gas kinematics, 2) a strong nuclear ionized outflow (blue and redshifted) in addition to the HG outflow.

\subsubsection{Stellar and gas kinematics around the E nucleus}


\begin{figure*}[t!]
    \centering
    \includegraphics[width = 0.9 \textwidth]{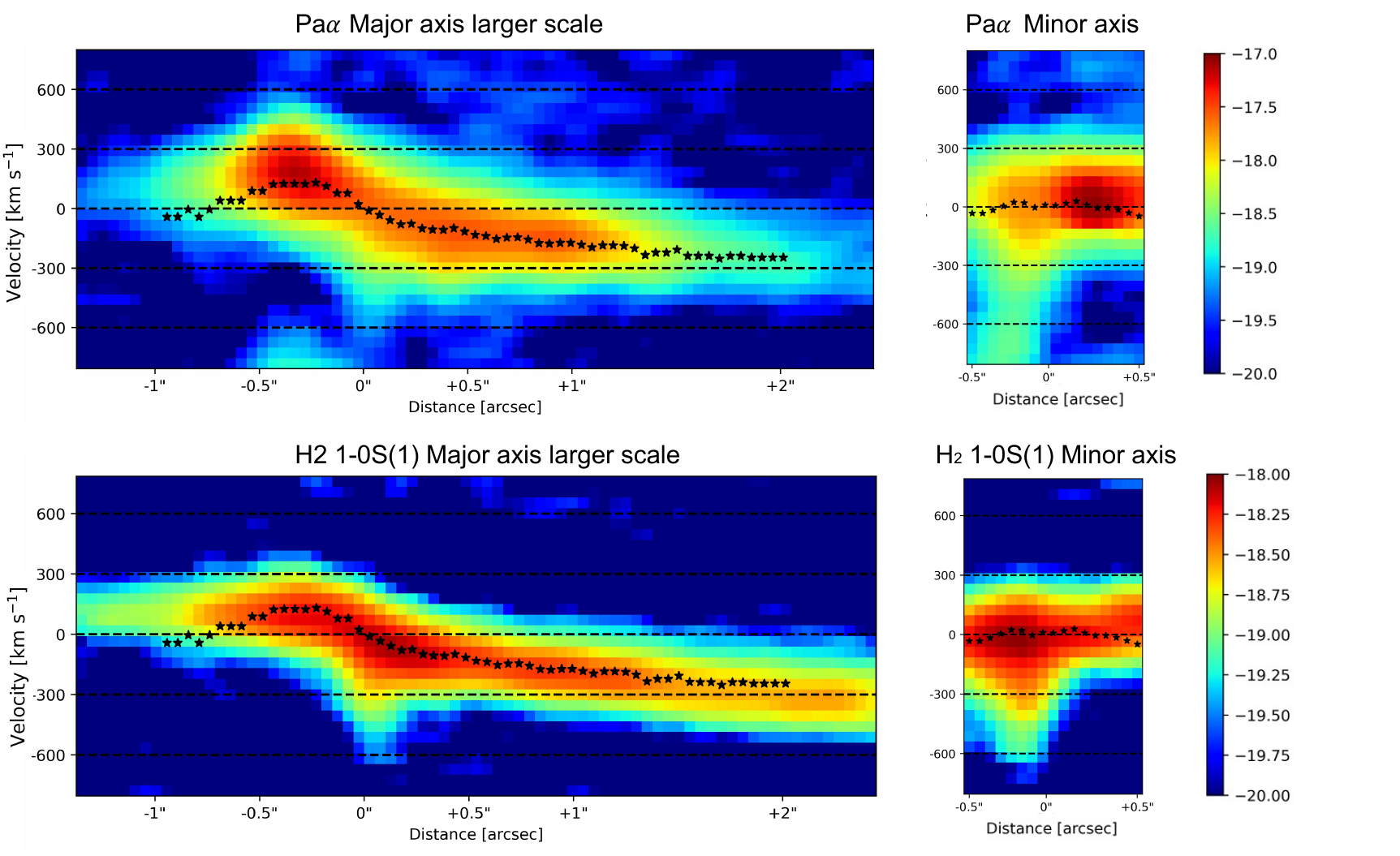}
    \caption{Position-velocity diagrams of large-scale rotational disk centered on the E nucleus. In the left panels, the positions vary along the mean kinematic major axis (PA = 45°), with positive distances toward south-west.
    In the right panels, the positions vary along the minor axis (PA = 135°), with positive distances toward north-west. The upper panels show the PV diagrams of Pa$\alpha$, while the lower panels show PV diagrams of H$_{2}$ 1-0~S(1). Black stars identify the stellar kinematics. The zero-velocity refers to z$^{*}_{E}$.  }
    \label{fig:PVrotational disc}
\end{figure*}

Figure \ref{fig:PVrotational disc} shows the 
PV diagram along the major axis of the stellar rotational disk centered at the position of E nucleus (PA = 45°). 
Since it was not possible to decouple the contribution of the small-scale disk surrounding the E nucleus from the large-scale disk, the major axis of the small-scale disk was assumed to be aligned with the one of the large-scale disk computed in the previous subsection. We set the zero velocity at the redshift of the E nucleus ($z_{E}^{*}$). 

The stellar component follows the rotation of the gas around the nucleus. In the ionized gas (Fig. \ref{fig:PVrotational disc}, top), a peak in the flux is observed at redshifted velocities $\sim250$~\kms, representing a bright clump at the north of the E nucleus (NE cluster in \citealt{Perna2024}), not observed in the hot molecular phase.

A blueshifted nuclear outflow is  observed in  both the hot molecular gas (bottom) and the ionized gas (top), reaching velocities of about $-600~$\kms. This emission forms the vertex of the SE conical outflow we isolated in Sect.~\ref{sec_Ose}.





\section{ISM properties}
\label{sec:ISMproperties}
\subsection{Attenuation}
\label{Sec_Extinction}
We used the emission-line ratios \feii~1.644/1.257, Br$\gamma$/Pa$\beta$ and Pa$\alpha$/Pa$\beta$ as  attenuation diagnostics. We measured the line ratios from each spaxel, and scaled them to calculate $E(B-V)$ as  follows (\citealt{Dominguez2013}):
\begin{equation}
E(B-V) = \frac{2.5}{k(\lambda_\mathrm{L1})-k(\lambda_\mathrm{L2})}\log_{10}\left(\frac{\mathrm{(L1/L2)_{obs}}}{\mathrm{(L1/L2)_{in}}}\right)
\label{ext}
\end{equation}

where L1 and L2 are the lines used for the diagnostic, and k$(\lambda_\mathrm{L2})$ and k$(\lambda_\mathrm{L1})$ are the values of the  reddening curve evaluated
at L1 and L2 wavelengths.
The intrinsic ratios are taken respectively from \cite{Bautista1998, Hummer1987} and from CLOUDY (\citealt{Ferland2017}) assuming  a temperature of 10$^{4}$~K and a density $n_{e} = 100$ cm$^{-3}$. We adopted the \cite{Calzetti2000} attenuation law valid between 0.12 to 2.2~$\mu$m with R$_{v}$ = 4.05 estimated from optical-IR observations of four starburst galaxies.

\begin{figure*}
    \centering
    \includegraphics[width = 1 \textwidth]{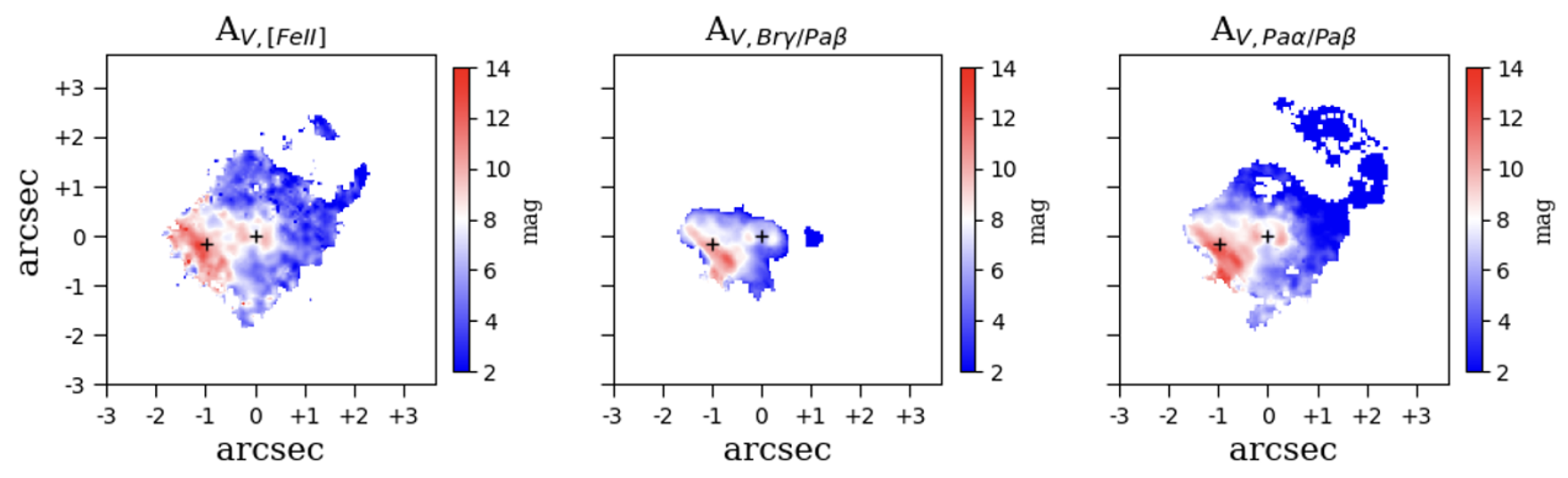}
    \caption{Dust attenuation maps from the\feii~1.257~$\mu$m to \feii~1.640~$\mu$m ratio (left), Br$\gamma$/Pa$\beta$ (middle) and Pa$\alpha$/Pa$\beta$ (right). 
}
    \label{Extinction}
\end{figure*}

Figure \ref{Extinction} shows the dust attenuation maps estimated from the different emission line ratios with a cut in S/N (>3) for each line, as A$_{V} = 4.05\times E(B-V)$. The attenuation A$_{V}$ varies between 2 and 14 mag, reaching higher values in the E nucleus (consistent with \citealt{Perna2024}) and lower values in the outer part of the FoV. The three attenuation maps computed with different line ratios are consistent. The attenuation computed from Eq.~\ref{ext} integrating the fluxes in the central region (where the S/N of the faintest line Br$\gamma$ > 3) varies less than 20\% between the three attenuation diagnostics.  These values are also consistent with the stellar attenuation computed by \cite{Engel+11} through the K band, A$_v = 6-12$,  while our values are a factor 2-3 higher than their estimates through Pa$\alpha$/Br$\gamma$.
\subsection{Star formation}

We derived the SFR from the Pa$\alpha$ dust corrected luminosity by summing the contribution of the perturbed rotating discs as defined in Sect. \ref{Perturbed_Rotating_Disc}, excluding all the outflow components.
We used the Pa$\alpha$ flux corrected for the dust attenuation spaxel per spaxel derived from the Pa$\alpha$/Pa$\beta$, resulting in an increase of the Pa$\alpha$ flux of a factor $\sim$2.2, and we followed \cite{Reddy+23}:
\begin{equation}
    \mathrm{SFR}(\mathrm{Pa\alpha})\left[M_{\odot} yr^{-1}\right] = \mathrm{C(Pa\alpha)} \times \mathrm{L(Pa\alpha)}\left[erg s^{-1}\right]
\end{equation}
where C(Pa$\alpha$) = 3.90$\times10^{-41}$ assuming an upper mass cutoff of the IMF of 100 M$_{\odot}$ and a solar metallicity. We derived a total SFR of $\sim$ 8~M$_{\odot}$ yr$^{-1}$. 
The value of SFR found with the recombination line is significantly lower than ones computed  in a similar area with FIR and radio data ($\sim$ 200~M$_{\odot}$~yr$^{-1}$, from \citealt{Varenius2016} and \citealt{Pereira2021}), possibly indicating that a large fraction of the SF is totally blocked by dust at NIR wavelengths (\citealt{Gimenez2022, Perna2024}). 

The largest contribution to the emission of recombination lines comes from the elongated region located between the two nuclei with relatively faint H$_2$ emission. In Fig. \ref{fig:PaH2_SFRdensity}a we show the ratio of the total Pa$\alpha$ flux to the total H$_{2}$ flux, which peaks up to 30 north of the E nucleus and drops to values less than one in the outermost regions. We also observed blobs in which the ratio reaches values of approximately two (in log: 0.3-0.4 from Fig.\ref{fig:PaH2_SFRdensity}a); one of the largest is located about 1" west of the W nucleus. These blobs lie along a preferential NS direction, probably along a filament or dust lane where SF is more intense.  

Figure \ref{fig:PaH2_SFRdensity}b shows the SFR surface density map, $\Sigma_{SFR}$. The $\Sigma_{SFR}$ is highest in the elongated region between the two nuclei (10--40~M$_{\odot}$ yr$^{-1}$ kpc$^{-2}$), while the other regions have value < 1~M$_{\odot}$~yr$^{-1}$ kpc$^{-2}$.  The $\Sigma_{SFR}$ calculated over the entire FoV ($\sim$ 1.5 kpc $\times$ 1.5 kpc) has a value of $\sim$3 M$_{\odot}$ yr$^{-1}$ kpc$^{-2}$. Similar values are reported in a sample of ULIRGs studied with VLT/SINFONI (\citealt{piqueras_lopez_2012}), ranging between 0.4 and 2~M$_{\odot}$~yr$^{-1}$ kpc$^{-2}$. Based on FIR measurements, the SFR surface densities of the nuclear region of Arp 220 reaches values higher than 10$^3$~M$_{\odot}$~yr$^{-1}$ kpc$^{-2}$. As discussed in \cite{Perna2024}, this discrepancy may be explained taking into account different factors: the high obscuration caused by the dust, the different scale times of SF that the two bands (FIR and NIR) trace and to the possible presence of AGN that is completely obscured in the NIR. Typical star formation surface density values computed in radio band in ULIRGs are $50 - 1000$~M$_{\odot}$~yr$^{-1}$~kpc$^{-2}$ in the nuclear region ($<1$~kpc$^{2}$), and $5 -500$~M$_{\odot}$~yr$^{-1}$~kpc$^{-2}$ on scales of few kiloparsec \citep{Lucatelli2024}.


\begin{table}[t!]
\tabcolsep 3.pt 
\centering
\caption{
Properties of the high-velocity molecular and ionized outflows.} 

\begin{tabular}{lcccc}

\multicolumn{5} {c} {General properties of the Nuclei} \\
\hline
   & E Nucleus &  & W Nucleus &  \\

\hline
SFR(radio/FIR)$^{(1)}$  & 60 – 80 &   & 110-150  &  \\
SFR(Pa$\alpha$)$^{(2)}$ & 0.17 &  & 0.52  &  \\
L$_{AGN}^{IR}$$^{(3)}$ [erg s$^{-1}$] & $1.1\times 10^{44}$ &  & $2.5\times 10^{44}$ & \\
L$_{AGN}^{X}$$^{(4)}$ [erg s$^{-1}$] & $8.3 \times 10^{43}$ &   & $2.5 \times 10^{43}$ & \\
Hot-to-cold$^{(5)}$ & $3\times 10^{-4}$ & & $10^{-4}$ & 

\\
 \hline
 \\
\multicolumn{5} {c} {General properties of the outflows}
\\
\hline
   & SEO & HGO & NWB & WNO \\

\hline
$R_\mathrm{out}^{(6)}$ [kpc] & 0.4 & 0.6 & 1 & 0.15 \\
$n_\mathrm{e}^{(7)}$  [10$^{3}$cm$^{-3}$] & <17 &  3.8$_{-1.3}^{+2.2}$ & < 1.8 & 5$_{-2}^{+2.5}$ \\
$v_\mathrm{out}^{(8)}$ [km s$^{-1}$]& 600$\pm$150 & 400$\pm$140 & 600$\pm$180 & 400$\pm$70 \\
t$_\mathrm{dyn}^{(9)}$ [Myr] &  0.65 & 1.5 & 1.7 & 0.42\\
$A_\mathrm{v}^{(10)}$ [mag] & 8.9 & 4.5 & 3 & 8 \\

 \hline

\\
\multicolumn{5} {c} {Hot Molecular Gas} \\
\hline

\hline
$ M_\mathrm{out}^{(11)}$ [10$^{4}$M$_\odot$] & 0.07 & 0.2 & 0.1 &  \\
$\dot M_\mathrm{out}$ [M$_\odot$ yr$^{-1}$] & 0.001 & 0.001 & 0.0006 & \\
${E}_\mathrm{out}^{(12)}$ [10$^{52}$ erg s$^{-1}$]   & 0.2 & 0.3 & 0.36 & \\
$\dot {E}_\mathrm{out}^{(13)}$ [10$^{40}$ erg s$^{-1}$]   & 0.009 & 0.006 & 0.007 & \\
\hline
\\
\multicolumn{5} {c} {Ionized gas, assuming \feii-based n$_e$} \\
\hline

\hline
$ M_\mathrm{out}$ [10$^{4}$M$_\odot$] & 0.5 & 4 & 6 & 9 \\
$\dot M_\mathrm{out}$ [M$_\odot$ yr$^{-1}$] & 0.007 & 0.025 & 0.03 & 0.26 \\
${E}_\mathrm{out}$ [10$^{52}$ erg]   & 1.7 & 6 & 18 & 16\\
$\dot {E}_\mathrm{out}$ [10$^{40}$ erg s$^{-1}$]   & 0.08 & 0.12 & 0.3  & 1.4 \\

\hline

\multicolumn{5} {c} {Ionized gas, assuming \sii-based n$_e$ = 170 cm$^{-3}$} \\
\hline

\hline
$ M_\mathrm{out}$ [10$^{4}$ M$_\odot$] & 50 & 90 & 60 & 270 \\
$\dot M_\mathrm{out}$ [M$_\odot$ yr$^{-1}$] & 0.7 & 0.6 & 0.4 & 8 \\
${E}_\mathrm{out}$ [10$^{52}$ erg]   & 170 & 130 & 190 & 460\\
$\dot {E}_\mathrm{out}$ [10$^{40}$ erg s$^{-1}$]   & 8 & 2.7 & 4  & 40 \\

\hline
\hline
\\
\multicolumn{5} {c} {Total (ionized, hot and cold molecular)} \\
\hline
\hline

$ M_\mathrm{out}$ [10$^{4}$M$_\odot$] & 400 & 2090 & 560 & 270 \\
$\dot M_\mathrm{out}$ [M$_\odot$ yr$^{-1}$] & 5.7 & 12  & 3.2 & 8\\
${E}_\mathrm{out}$ [10$^{52}$ erg]   & 1170 & 3570 & 1940 & 460\\
$\dot {E}_\mathrm{out}$ [10$^{40}$ erg s$^{-1}$]   & 53 & 71 & 39  & 40 \\
$\mu^{(14)}$  & 0.08 & 0.09  & 0.05 & 0.06\\
$\dot {E}_\mathrm{out}$/L$_{AGN}^{IR}$ & 5$\times 10^{-3}$ & 3$\times 10^{-3}$ & 4$\times 10^{-3}$ & 2$\times 10^{-3}$ \\
$\dot {E}_\mathrm{out}$/L$_\mathrm{AGN}^\mathrm{X}$ & 6$\times 10^{-3}$ & 0.03 & 5$\times 10^{-3}$ & 0.02 \\
$\dot {E}_\mathrm{out}$/$\dot{E}_\mathrm{SFR}^\mathrm{radio/far-IR}$ & 0.009 & 0.007 & 0.006 & 0.004 \\

\hline

\end{tabular} 

Notes: (1) and (2) SFR computed from radio/FIR by \cite{Varenius2016} and \cite{Pereira2021}, and 
from Pa$\alpha$ by \cite{Perna2024}, respectively. (3) AGN luminosity from IR. (4) AGN luminosity from X-rays by \cite{Paggi2017}. (5) Hot-to-cold ratio. (6) Outflow radius. (7) \feii-based electron density. (8) Outflow velocity, calculated as the flux-weighted average of the velocity (defined as in Eq.~\ref{eq:vout}) of each spaxel and the associated errors. (9) Dynamical time. (10) Average attenuation. (11) Outflow mass. (12) Outflow kinetic energy. (13) Outflow kinetic power. (14) Outflow mass loading factor.

The AGN luminosity refers to the nucleus from which the outflow originates. The SEO and NWB are attributed to the E nucleus, while the HGO and WNO to the W nucleus.  
\label{tab:kinematics}
\end{table}

\subsection{Outflows properties and energetics}
\label{subsec:outflow_properties}
In this section, we estimate the outflow energetics of the molecular and of the ionized gas.
To estimate the mass of the outflowing gas in the hot molecular phase, we followed \cite{Scoville+82} and \cite{Mazzalay+13}:

\begin{equation}
     M_\mathrm{out,\ H_2} = 5.1 \times 10^{13} \left(\frac{D_\mathrm{L}}{\mathrm{Mpc}}\right)^{2}\left(\frac{F_\mathrm{1-0~S(1)}}{\mathrm{erg~s^{-1}cm^{-2}}}\right)10^{0.4\mathrm{A_{2.2\mu m}}} M_{\odot}
     \label{Scoville}
\end{equation}
where $M_\mathrm{out,\ H_2}$ is the hot molecular mass traced by H$_{2}$, $D_{L}$ is the luminosity distance, and $A_{2.2\mu\text{m}}$  is the  attenuation at 2.2~$\mu$m. 
This equation assumes thermalized gas conditions and $T = 2000$~K, with a population fraction in the ($\nu$, J) = (1, 3) level of $f(1,3) = 0.0122$. We applied Eq. \ref{Scoville} for each spaxel and then we summed over the region of the spatially resolved outflows (namely, the SEO, the HGO, the NWB, and the WNO) to compute the total masses. 
We computed $A_{2.2\mu m}$ as shown in Sect.~\ref{Sec_Extinction} through Pa$\alpha$/Pa$\beta$, since it involves the lines with higher S/N among the line ratios considered. However, when computing the properties of the outflow in the spaxels where we could not infer the  attenuation due to low S/N <3 in Pa$\alpha$ and Pa$\beta$, 
we assumed A$_{V}$ corresponding to the mean value in the defined region. 

To compute the ionized outflow mass, we adopted the simplified model by \cite{Genzel2011}, valid for case B recombination of fully ionized gas with $T\sim10^{4}$ K. The mass of outflowing gas is given by
\begin{equation}
     M_\mathrm{out,\ ion}/M_{\odot} = 3.2 \times 10^{5} \left(\frac{L_\mathrm{H\alpha}}{10^{40}~\mathrm{erg~s^{-1}}}\right)\left(\frac{n_{e}}{100~\mathrm{cm^{-3}}}\right)^{-1}
     \label{Cresci 2017_f}
\end{equation}
where L$_{H\alpha}$ is derived from L$_{Pa\alpha}$ corrected for attenuation assuming a theoretical ratio Pa${\alpha}$/H${\alpha}$ = 0.11, and $n_{e}$ is the electron density. 

The determination of the ionized gas mass is significantly influenced by $n_{e}$, which can span a few orders of magnitude, from 10$^{2}$ to 10$^{4}$cm$^{-3}$ (e.g., \citealt{Villar2014, Perna2017b, Schreiber2019, Isobe2023}). 
We derived the electron density through the line ratio \feii 1.677~$\mu$m/\feii 1.644~$\mu$m. We fitted the integrated spectrum around the lines of interest 
considering in the model also the Br11 line, since it is blended with \feii 1.677~$\mu$m. Since the latter line is very faint at spaxel level, we integrated the spectra over the individual outflow regions SEO (Fig. \ref{SE_Outflow}), HGO (Fig. \ref{Hourglass}), NWB (Fig. \ref{Northern_Bubble}), and WNO (Fig. \ref{rotational disc}, bottom panels), to infer average electron densities for each outflow. Moreover, due to the low S/N of \feii1.677, we considered the total profile of \feii 1.677 $\mu$m and 1.644 $\mu$m, without distinguishing  the contribution of non-outflowing components. 
We obtained electron densities in the range log (n$_{e}$/cm$^{-3}$) $\approx 3-4$ from the relation in \cite{Koo2016}.

We found that the region with the highest electron density is the WNO (n$_e=~5000_{-2000}^{+2500}$ cm$^{-3}$). For the HGO, we found n$_e = 3800_{-1300}^{+2200}$~cm$^{-3}$. We also integrated the spectrum and computed the electron density in both the red- and blueshifted region of the HG outflow, separately. We found n$_e = 1600_{-500}^{+800}$~cm$^{-3}$  for the redshifted and a 3$\sigma$ upper limit  of 5$\times 10^4$ cm$^{-3}$ in the blueshifted part. In the SEO and in the NWB the emission of \feii~1.677~$\mu$m is not detected, resulting in upper limits on the electron density: 
$< 1.7 \times 10^{4}$ cm$^{-3}$, and 
$< 1800$~cm$^{-3}$, respectively. 
All the inferred values are reported in Table~\ref{tab:kinematics}. 

The $n_{e}$ estimates through the \feii transitions are higher (a factor $\sim$ 10-100) than those found from the \sii doublet in the optical band with MUSE (170 cm$^{-3}$, \citealt{Perna2020}). This may indicate a gas stratification in the nuclear regions of Arp 220, with \feii emission, having higher critical density (N$_c$ $\sim 10^{4}$ cm$^{-3}$), tracing the densest part of the ISM, and \sii (N$_c$ $\sim 10^{3}$ cm$^{-3}$) tracing the more diffuse gas. It is worth noting that using a constant density of $n_{e}$ $\sim$ 10$^2$ cm$^{-3}$, as the one derived by \cite{Perna2020} using \sii, could lead to an overestimation of the gas masses and the energetics,  because this line emission likely traces partially ionized gas  with lower densities (\citealt{Davis2011,Revalski2022}).  On the other side, our estimation of the $n_{e}$ from \feii doublet comes only from the nuclear (WNO) and the most compressed regions (HGO), where the faint \feii~1.677~$\mu$m line is detected. Thus, this electron density may not be representative of the density in the full FoV.  Therefore, we chose to estimate the outflow properties with electron density estimations from both \feii and \sii that may represent reasonable lower and upper boundaries. 



For each isolated outflow in Sect. \ref{sec:separations}, we determined the mass outflow rate, $\dot M_{out}$, at a given radius $R_{out}$ by using the assumptions of a constant outflow speed $v_{out}$ and spherical (or multiconical) geometry. Following \cite{Lutz2020}, we have

\begin{equation}
     \dot M_\mathrm{out} = 1.03 \times 10^{-9} \left(\frac{v_\mathrm{out}}{~\mathrm{km~s^{-1}}}\right)\left(\frac{M_{out}}{M_{\odot}}\right)\left(\frac{R_{out}}{kpc}\right)^{-1} \times \textit{C} \, M_{\odot}~ yr^{-1}
     \label{Lutz2020}
\end{equation}
where \textit{C} is a factor that depends on the outflow history. We adopted a constant mass outflow rate (C=1).
We calculated the radius from the flux maps of each outflow. In particular, we measured $R_{out}$ as the maximum extension from the corresponding nucleus. Specifically, for the HGO and WNO, we calculated the distances from the W  nucleus, while for the SEO and NWB we calculated them from the E nucleus. We associated the NWB to the E nucleus, in order to have a more conservative estimation of the mass outflow rate. This choice is also motivated by the fact that the outflows launched by the W nucleus are oriented along the NS direction (Sect. \ref{sec:HGoutflow}), while the one of the E nucleus is along the SE direction. Moreover, the 100~pc scale molecular outflow launched by the E nucleus and traced by CO \citep{Wheeler2020} is similarly oriented along the SE-NW direction.

We used the definition of the outflow velocity in \cite{RV2013} as
\begin{equation}
    v_\mathrm{out} = |v_\mathrm{broad}| + 2\sigma_\mathrm{broad}
    \label{eq:vout},
\end{equation}
where $v_\mathrm{broad}$ and $\sigma_\mathrm{broad}$ are the velocities relative to the systemic velocity of the nucleus from which the outflow originates, and the velocity dispersion of the outflowing component, respectively. We computed the outflow velocities as the flux-weighted average of the outflow velocity of each spaxel. We assigned the standard deviation of the values in that region as the uncertainty on v$_\mathrm{out}$.
The properties of the ionized and molecular gas outflows for the analyzed regions are reported in Table~\ref{tab:kinematics}.

The outflow masses and the mass outflow rates of the hot molecular gas  vary from 700~M$_{\odot}$ in the SEO to 2000~M$_{\odot}$ in the HGO, and from $6\times10^{-4}$~M$_{\odot}$~yr$^{-1}$ for the NWB to $1\times10^{-2}$~M$_{\odot}$ yr$^{-1}$ for the HGO, respectively. 
For the ionized gas, the outflow mass varies from a minimum of $5\times10^{3}$ M$_{\odot}$ for the SE outflow to a maximum of $9\times10^{4}$ M$_{\odot}$ fort the WNO, calculated with the density from the \feii doublet.  The highest mass outflow rate of ionized gas is measured in the WNO with a value of 8~M$_{\odot}$~yr$^{-1}$, estimated using the density from the \sii, while the lowest is measured in the NWB with a value of 0.4~M$_{\odot}$~yr$^{-1}$.

We also computed the kinetic energy and the kinetic power of the outflow as
$E_\mathrm{out} = \frac{1}{2} M_\mathrm{out}v_\mathrm{out}^{2} $ and  $    \dot E_\mathrm{out} = \frac{1}{2}\dot M_\mathrm{out}v_\mathrm{out}^{2}$, respectively.
We found that the contribution to the energetics of the ionized outflow is about 1-3 orders of magnitude higher than that of the hot molecular gas, depending on the density used in the computation. The kinetic energy of the ionized gas outflows ranges between 1.7$\times 10^{54}$ erg for the SEO to 4.6$\times 10^{54}$ erg for the WNO, estimated with \sii.  


\subsection{Three-dimensional outflow modeling }
\label{sec:modellingwithMOKA}
To measure the geometrical parameters of SEO, HGO and the NWB outflows we used the modeling framework called Modeling Outflows and Kinematics of AGN in 3D (\MOKA) presented in \cite{Marconcini2023}. By assuming a geometry and a simple constant radial velocity field, the model is able to retrieve the outflow inclination with respect to the line of sight (LOS), and the intrinsic outflow velocity. Full details and the implementations of the model are discussed in \cite{Marconcini2023}. Here we just show the results of the model for the hot molecular gas, by using a conical geometry for the SEO and two expanding bubbles for HGO and NWB, assuming a constant radial velocity (See Appendix B). We also assumed an intrinsic dispersion velocity of 30~\kms to account for the turbulent nature of the bubbles.  Since we found that the ionized gas has a similar kinematics, the results of the model for the ionized phase would be the same. We did not attempt to model the WNO since it is very compact and we could not establish a clear geometry to be assumed.

\begin{figure}[t!]
    \centering
    \includegraphics[width = 0.5\textwidth]{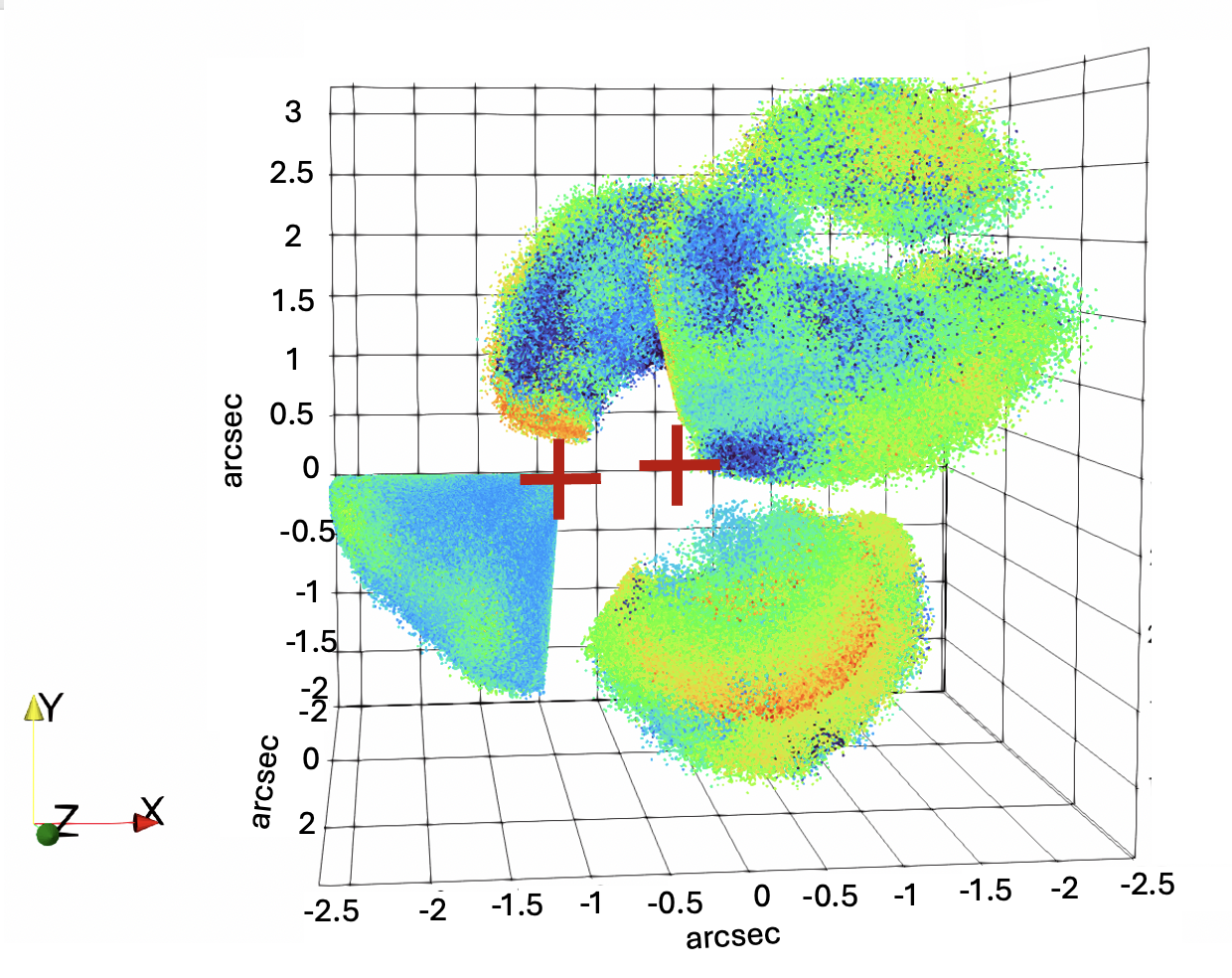}
    \caption{Three-dimensional representation of the hot molecular outflow structures detected in Arp~220 modeled with MOKA3D. The XY represents the plane of the sky, while the Z axis is the LOS. The color intensity represents the intrinsic cloud emission, bluer colors represent lower emission clouds, while redder colors represent high-emission clouds. (See \cite{Marconcini2023} for the details of the model.) }  
    \label{fig:MOKA3D}
\end{figure}

\begin{figure*}[t!]
    \centering
    \includegraphics[width = 0.95\textwidth]{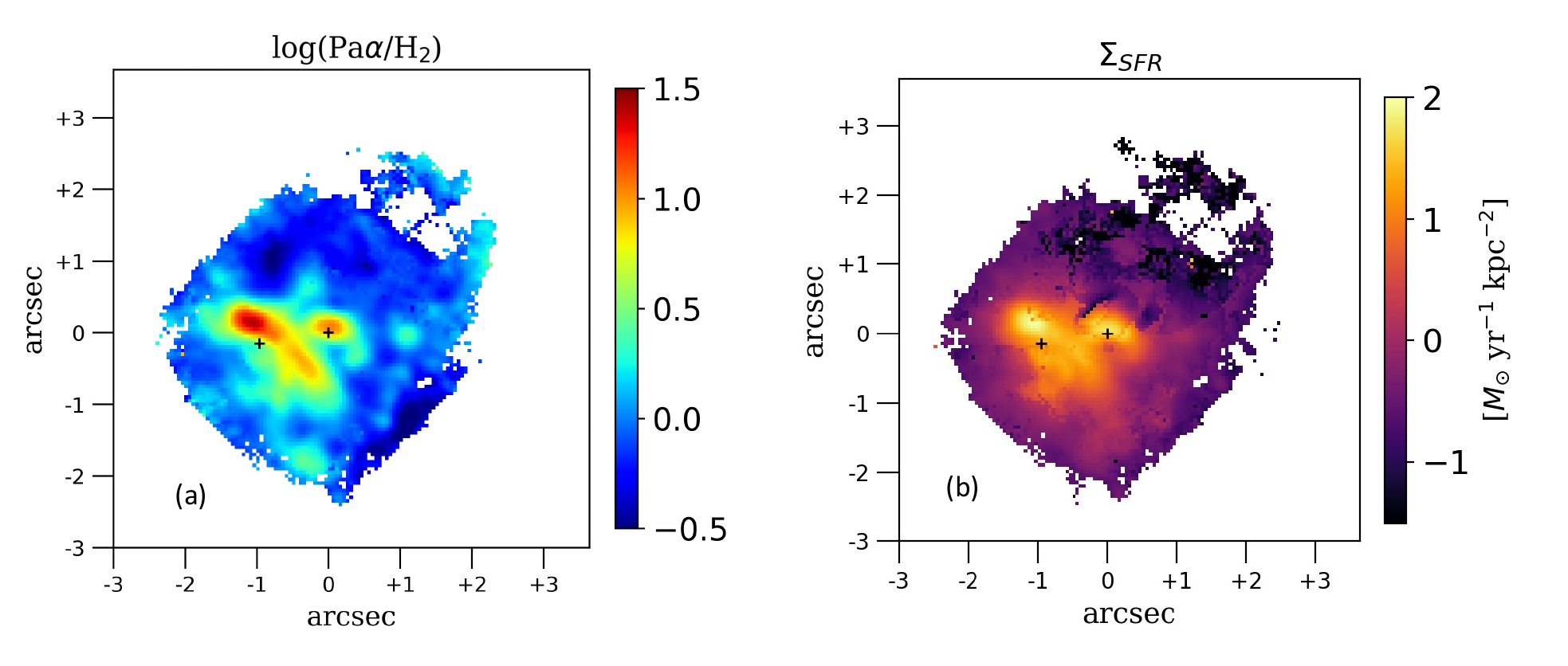}
    \caption{Maps of ionized over hot molecular gas ratio and SF surface density. (a) Spatially resolved flux ratio log(Pa$\alpha$/H$_{2}$) computed from the total profile of the two lines. (b): Surface density map of the SFR inferred from Pa$\alpha$ after subtraction of the outflowing components.}
    \label{fig:PaH2_SFRdensity}
\end{figure*}

In Fig. \ref{fig:MOKA3D} we show the 3D model of the outflows detected in the nuclear region of Arp~220. The geometry parameters derived from \MOKA\ are as follows. 
\begin{itemize}
\item The aperture of the SEO is 80° with an inclination respect to the LOS of 70° and PA of 230° measured clockwise from north. This is consistent with a conical outflow launched perpendicular from the E circumnuclear disk having an almost edge-on orientation, according to \citet{Wheeler2020}.
The velocity retrieved for the SEO by \MOKA is $\sim$700~\kms, consistent within $<1\sigma$ with the velocities computed as defined as in Eq.~\ref{eq:vout}), but slightly higher.
\item  The HGO has an inclination of --60° with respect to the LOS and a velocity of $\sim$400 \kms for the blueshifted side, and 120° and $\sim$600 \kms for the redshifted side. The discrepancies between the velocity in the redshifted and blueshifted side could be explained by the possible ISM porosity and hence different propagation of the bubbles. The inclination of the W circumnuclear disk is $\sim$120° with respect to the LOS (\citealt{Wheeler2020}), which mean that the axis of the bubble is not exactly perpendicular to the disk, but has an offset of about 30°.
\item For the NWB we found an inclination of 68° with respect to the LOS and a velocity of $\sim$750~\kms, also in this case slightly higher than our flux-weighted velocity measurement. 
\end{itemize}
In all outflows, the velocities defined by Eq.~\ref{eq:vout} seem to underestimate the velocities found by the \MOKA\ even though they are consistent within 1-2$\sigma$. This is due to the fact that the model takes into account the intrinsic velocity of the outflow corrected for all projection effects. In conclusion, the kinematic outflow properties inferred with \MOKA and the method outlined in Sect.\ref{subsec:outflow_properties} are consistent within the uncertainties.

\section{Discussion}
\label{sec:Discussion}



The ubiquity of multiphase outflows in ULIRGs was established through spatially resolved observations in the visible band, thanks to integral field instruments like VIMOS and MUSE (\citealt{Arribas+14,Westmoquette2012, Venturi2021, Musiimenta2024}), in the near IR band with Gemini/NIFS, Keck/OSIRIS, and VLT/SINFONI (\citealt{Engel2011, Riffel2015, Storchi-Bergmann2010,Pereira-Santaella2016,Emonts2017, U2019, Schonell2019}), and in the radio band with ALMA (\citealt{Pereira-Santaella2016}).

Simulations show that outflows likely contain cold, dense clouds that are accelerated by a stream of warm and hot gas (e.g., \citealt{Wagner2012, Zubovas2014}). However, the direct comparison of outflows identified with different tracers (and facilities), required to test such predictions, has been so far limited to very few targets. A key challenge in studying the direct impact of outflows on the ISM is the need for both high sensitivity and spatial resolution. 
Ground-based near-IR integral field facilities like SINFONI, OSIRIS and NIFS 
mainly traced the hot molecular (e.g., H$_{2}$) and ionized (Br$\gamma$) gas with high spatial and spectral resolution. However, due to limitations in sensitivity and the presence of airglow emission lines, it has not always possible to fully characterize these outflows and compute directly their properties  (e.g \citealt{Riffel2014, Emonts2017, U2019, Bianchin2022}). 
This is critical in order to characterize the connection between the warm molecular and the ionized gas outflow with the cold molecular gas. In fact, thanks to the advent of millimeter/submillimeter telescopes such as ALMA, high-spatial-resolution images of cold molecular gas in nearby galaxies have enabled us to reveal the structure and the morphology of starburst-driven and AGN-driven molecular outflows in high details (e.g \citealt{Salak2020, Lamperti2022, Holden2024}).  


\subsection{The multiphase outflows}
Cold molecular outflows have been detected in the nuclear region of Arp 220 with different emission line tracers, such as HCO$^+$, H$_2$O, H$_2$O$^+$, OH, and OH$^+$ (\citealt{Sakamoto2009,GonzalezAlfonso2012, Veilleux+13, Tunnard2015, Martin2016, Zschaechner2016}). However, most of them are unresolved and therefore their morphology could not be constrained. Spatially resolved outflows studies have been conducted by \cite{Barcos-Munoz2018} and \cite{Wheeler2020} with ALMA, and \cite{Perna2020} with MUSE. Thanks to the possibility of resolving multiphase outflows with NIRSpec, we can compare the outflow properties of hot molecular and ionized phases derived in this work, with those obtained from previous observations for cold molecular gas and ionized gas at larger scale. This will help us understand the connection between multiphase and multiscale outflows. 
After discussing the structures of these outflows, we examine the mechanisms responsible for launching them.

\subsubsection{W nucleus-driven outflows}\label{sec:discussionWnuc}

\cite{Barcos2018} reported a cold molecular outflow launched by the W nucleus, traced by HCN(1-0), extending for 120 pc, with a deprojected velocity of 850 \kms\ and a lower limit 
of the H$_2$ mass for the redshifted and blueshifted component of $\gtrsim 8\times 10^{6} M_{\odot}$ and $\gtrsim 2\times 10^{6} M_{\odot}$, respectively. These lower limits consider the possible presence of additional outflowing material at lower velocities, not accounted due to its overlap with the disk emission.  \cite{Wheeler2020} found an upper limit for the total H$_{2}$ molecular mass of $\lesssim 2\times 10^{7} M_{\odot}$ for the redshifted outflow, and $2\times 10^{6} M_{\odot}$ for the blueshifted, traced by CO(3--2) gas. The upper limit here accounts the possible contamination of H$^{13}$CN (4--3) emission, very close to the CO(3--2) transition.  

We detected an HG-shape outflow originating from the W nucleus extending for 600 pc, with a hot H$_2$ outflowing gas mass of 2000~M$_{\odot}$. The redshifted region accounts for $\sim$ 1200~M$_{\odot}$, while the blueshifted for $\sim$800~M$_{\odot}$. This outflow is in the same direction as the cold molecular outflow detected in \cite{Wheeler2020}, with receding gas northward and approaching gas southward. Therefore, we can reasonably assume that cold and hot molecular outflows are connected and associated with the same outflow event. Under this assumption, the total mass of the hot and cold molecular gas result in an estimated hot-to-cold molecular gas ratio of $\approx 10^{-4}$, obtained considering both receding and approaching gas.
This order-of-magnitude estimate is consistent 
with the values $1-7\times 10^{-5}$ found  in nearby ULIRGs \citep{Pereira-Santaella2016,Emonts2017,Pereira2018, Lamperti2022}.


%
After comparing the total outflowing mass of ionized and total molecular gas in the W nucleus and hence accounting for both HGO and WNO, we report a mass outflow ratio between the two phases of 0.16 assuming \sii-based electron densities ($\sim 6 \times 10^{-3}$ with the density estimated with \feii). This is not unusual since the molecular phase usually dominates the mass outflow rate budget, exceeding the ionized gas phase by one to three orders of magnitude (\citealt{RV2013, Burillo2015, Carniani2015, Fiore2017, Cresci2023}). 

\cite{Perna2020} identified an extended structure associated with  high-velocity ionized gas (see their Fig. 19, velocity channel 125 \kms < |v| < 375\kms) from northeast to southwest, that appears to be aligned with the hourglass-shaped outflow launched from the W nucleus. This  suggests that the HGO could even be connected to the outflow detected at larger scales. 
Alternatively, 
they could trace different ejection events along the same direction.

\subsubsection{E nucleus-driven outflows}\label{sec:discussionEnuc}
\cite{Wheeler2020} reported a collimated and extended molecular outflow in the E nucleus traced by CO transitions (J$=~3\rightarrow~2$).
The outflow found in the present work (SEO), launched from the E nucleus, is consistent with the direction of the outflow observed in the cold molecular gas, blueshifted toward the south direction. 
\cite{Wheeler2020} reported a mass for the blueshifted outflow of 2.7$\times 10^{6}$ M$_{\odot}$, which, under the assumption that the CO outflow is associated with the SE outflow detected in our data, would lead to a hot-to-cold molecular gas ratio of $\sim3\times 10^{-4}$. 
This is compatible with the value we inferred in Sect.~\ref{sec:discussionWnuc} for the outflow launched by the W nucleus.

We attributed the NW bubble to the E nucleus, since this structure could represent the counterpart of the blushifted outflow detected in the SE direction. However, unlike the HGO, in this case there is no clear blushifted and redshifted part of the outflow. This may be due to different gas ejection events, or to the fact that the redshifted part of the outflow is more obscured. Another possibility is that we included part of the redshifted outflow launched from the E nucleus in the HGO, since the two outflows may overlap, as we can see in Fig. \ref{Mask_Features}c. From Fig. \ref{fig:MOKA3D} we see that \MOKA assigns a higher flux weight to the clouds of the HGO near the W nucleus. This suggests the possibility of contamination from the redshifted counterpart of the SEO.


We find that the mass of the ionized gas in the NW bubble, assuming n$_e$ = 170 cm$^{-3}$,  is $\sim 6\times 10^{5}$ M$_{\odot}$, with a total energy of  $\sim 2\times 10^{54}$ erg.
\cite{Lockhart2015} estimated the properties of the NW bubble using H$\alpha$+[NII] narrow band imaging obtained with HST/WC3 with a resolution of 0.0396"/pixel, corresponding to 15 pc.
They used  assumptions different from ours on the density, on velocity of the bubble and also on the geometry. In fact, they adopted n$_{e}$ = 10 cm$^{-3}$, about one order of magnitude lower than that used in this work, a velocity of 250~\kms, around a factor 2 lower and a shell-like geometry.
Despite these different assumptions, their derived outflow mass ($3.2\times 10^{6}$ M$_{\odot}$) and energy ($2.0\times 10^{54}$ erg) are in agreement, within a factor of two, with our results.

We also compared the outflows detected with NIRSpec with the larger scale outflows observed with MUSE in \cite{Perna2020}. 
In that case, the most extreme velocities for the ionized outflow are observed along PA = 138° \footnote{Measured anti-clockwise from the north.}; these extreme motions are aligned with X-ray emission and with the minor axis of the stellar disk. 
The SEO and the NWB are found along the same PA, but at < 1 kpc scales, which might suggest that they are related to the kiloparsec-scale atomic outflow presented in \cite{Perna2020}. The ionized, large-scale bubble observed in MUSE ($\sim10$kpc) and the NW bubble could be, also in this case, the results of different outflow episodes. These multiple bubbles at different distances are very similar to those predicted by the TNG50 cosmological simulation as a result of AGN feedback (\citealt{Nelson2019}) and to those observed in  NGC 1275 (e.g., \citealt{Fabian2003, Sanders2007}) and in the Tea Cup (\citealt{Venturi2023}).


\subsection{ The launching mechanism } 
The presence of AGN in the Arp 220 nuclei has been investigated in many previous works (e.g., \citealt{Teng2015, Paggi2017, Yoast2019, Perna2024}), but conclusive evidence confirming their existence remains elusive. 
Furthermore, understanding whether outflows are powered by AGN or starburst presents an even greater challenge; in this section we relate the energetics of the Arp 220 multiphase outflows to the properties of AGN (if present) and SF, and we discuss possible scenarios.


To evaluate the effect of the outflows on the SF activity, we can compare their energetics with the SFR. However, the SFR varies depending on the band in which it is calculated, spanning values from 150 M$_{\odot}$~yr$^{-1}$ in the radio band (\citealt{Varenius2016}) to 0.5 M$_{\odot}$~yr$^{-1}$ through Pa$\alpha$ (\citealt{Perna2024}) in the W nucleus, and from 80 to 0.17 M$_{\odot}$~yr$^{-1}$  in the E nucleus. 
Since the SFR measurements obtained from optical and near-IR features 
may be  significantly affected by dust attenuation, leading to underestimation, we focus on the case where the SFR is calculated using the radio band.

Summing all the contributions of the outflows in the ionized and molecular phases in each nucleus, we derive a mass loading factor $\eta$ of $\sim$ 0.13 and 0.15 and a $\dot E_{out}$ of $1.1\times 10^{42}$ erg s$^{-1}$ and $9\times 10^{41}$ erg s$^{-1}$ for the W and E nuclei, respectively. We note that these values are only marginally  affected by the densities used, since the main contribution comes from the cold molecular gas.
These values are similar to those measured in other local ULIRGs (\citealp{Arribas+14, Chisholm2017, Lamperti2022}) and high-z star-forming galaxies (\citealt{Genzel2014,Heckman2015, Newman2012, Perna2017b, Perna2018}). Moreover, they are in agreement with models  where feedback from supernovae is the main outflow driver (e.g., \citealt{Finlator2008,Heckman2015}).

Following \cite{Veilleux2005}, we can derive the expected energy outflow rate by assuming proportionality with SFR (measured from radio band) and assuming solar metallicity. We derive  $\dot{E}_{SFR} = 1\times 10^{44}$ erg s$^{-1}$  for the W nucleus and $6\times 10^{43}$ erg s$^{-1}$for the E nucleus. Comparing these values expected for stellar activities with those obtained in previous paragraph, we find $\dot E_{out}/\dot E_{SFR}$ of the order of a few percent, 
in agreement with a starburst-driven outflow expectations. 


\cite{Paggi2017} estimated AGN luminosities in the $2-10$~keV X-ray band of $0.5 \times 10^{42}$ erg s$^{-1}$ and $1.6 \times 10^{42}$ erg s$^{-1}$ for the W and E nuclei that lead to bolometric luminosities of $2.5 \times 10^{43}$ erg s$^{-1}$ and $8.3 \times 10^{43}$ erg s$^{-1}$, respectively.
We also estimated the AGN luminosity from the total IR luminosity, assuming an AGN contribution of 6\% to  L$_{bol}$
(\citealt{Perna2021}) and using the relative contribution of each nucleus measured in the ALMA continuum maps by \citet{Pereira2021}: L$_{bol} = 2.5\times 10^{44}$ erg s$^{-1}$ (W) and $1.1\times 10^{44}$ erg s$^{-1}$ (E). 
It is worth noting that the AGN bolometric luminosity of the W nucleus inferred from FIR measurements is approximately 1~dex higher than the X-ray-based estimate, highlighting the challenges in constraining AGN activity in Arp 220. 
The outflow energetics computed in this section result in kinetic coupling efficiency values $\dot E_{out}/L_{AGN} = 0.005-0.05$ for the W nucleus, and $\dot E_{out}/L_{AGN} = 0.009-0.01$ for the W nucleus. 
These values are consistent with both theoretical predictions of kinetic coupling efficiencies in energy-conserving ($\sim$ 0.005–0.05, e.g., \citealt{Hopkins2010, Zubovas2012, Costa2014, KingPounds2015}) and radiation pressure-driven outflows ($\sim$0.001–0.01, e.g., \citealt{Costa2018, Ishibashi2018}). 

These results show that there are no arguments at present to prefer SF- to AGN-driven outflows, since both of the scenario are possible.
We note, however, that the typical dynamical timescales t$_{dyn}$ = R$_{out}$/$v_{out}$ associated with the outflows in Arp 220 are approximately 1 Myr (see Table \ref{tab:kinematics}). In contrast, the SFR~$\approx 100$~M$_\odot$~yr$^{-1}$ used to compare outflow energetics with expectations for starburst-driven outflows reflects the SF history over a much longer period (up to 100~Myr; \citealt{Kennicutt2012}). Following on these arguments, we consider two scenarios. In the former, we assume that the SFR in the past few million years is $<< 100$ \Msunyr. This would be consistent with the SFR measured in the host from UV, optical and near-IR observations (e.g., \citealt{Chandar2023,  Perna2020, Perna2024, Gimenez2022}), as well as with the emission from supernovae and supernovae remnants at wavelengths from 2 to 18 cm which can be responsible for up to  20\% of the total radio emission at GHz frequencies (but see the detailed discussion in \citealt{Varenius2019}). 
In this framework, the mass loading factors and outflow energetics would imply the presence of AGN winds, as SF would not provide enough energy to explain the outflow properties ($\eta>40 $, $\dot E_{out}/\dot E_{SFR}>3$). In the second scenario, we assume the presence of extremely high attenuation at the position of the two nuclei, of $A_V > 100$: such attenuation is required to match the SFR(PAH 3.3~$\mu$m) $\sim $ 1 \Msunyr with measurements from FIR and GHz frequencies. 
With this assumption, both starburst- and AGN-driven winds can be responsible for the observed outflows, according to the outflow energetics computed in this work. 

In conclusion, further investigations are required to constrain the current SFR and the AGN activity in the nuclear regions of Arp 220. This information is essential for understanding the mechanisms driving the multiphase and multiscale outflows observed in this galaxy.

\section{Conclusions}
\label{Conclusions}
Using the exceptional JWST\slash NIRSpec facility, we investigated the complex multiphase structure and kinematics  in the nuclear region of the ULIRG merging system Arp 220. We spatially resolved the emission of ionized gas (Pa$\alpha$) and hot molecular gas (H$_2$ 1-0 S(1) 2.12\,$\mu$m) in the central kiloparsec of the nuclear region of Arp 220 in order to identify and disentangle several high-velocity multiphase outflows associated with the two nuclei of the system.  
We modeled the emission of the outflows  with \MOKA and inferred the intrinsic geometric and kinematic properties, finding agreement with the properties inferred from standard methods.
We calculated the outflow mass and energetics of both ionized and hot molecular gas, and compared them with theoretical expectations for outflows driven by SF and AGN activity. 
In the following, we summarize our main findings.

\begin{itemize}

\item
We detected outflowing H$_2$ and ionized gas with an hourglass shape launched by the W nucleus with an inclination of -60° with respect to the LOS as retrieved from \MOKA and extending out to a distance of 600 pc with velocities of $\sim$ 500 \kms. We also found an ionized compact outflow around the W nucleus, with velocities of up to 550 \kms and without a hot molecular gas counterpart. 
\item
We observed a hot molecular and ionized conical outflow in the southwest direction launched by the E nucleus with an inclination of 70° with respect to the LOS reaching velocities of up to 600 \kms and extending to the edges of the NIRSpec FoV (i.e., at least up to 400~pc). We also detected a multiphase blueshifted large-scale ($\sim$ 1 kpc) bubble in northwest direction with an inclination of 68° with respect to the LOS and with velocities of up to 1100 \kms. These high-velocity components are aligned with the larger-scale ($\sim 10$~kpc) outflow observed with MUSE, suggesting that they are part of the same outflow event originating from the E nucleus. 

\item 
We combined NIRSpec and ALMA outcomes to infer a hot-to-cold molecular gas ratio of $\sim 10^{-4}$ 
for the outflows launched by the W and E nuclei, which is consistent with previous measurements for other ULIRGs from the literature. 
Considering the material launched by the E and W nuclei together, 
the largest contribution to the total outflow mass rate 
comes from cold molecular gas ($\sim~20$~\Msunyr), followed by ionized gas ($\sim~10$~\Msunyr, assuming \sii-based n$_e$) and hot molecular gas ($\sim$ 0.003~\Msunyr). 
\item 
We discussed the possible origin of the outflows, and we conclude that there are no arguments at present to prefer SF- over AGN-driven outflows  since both scenarios are possible. In particular, both recent nuclear activity with SFR $\sim 100~$\Msunyr and AGN luminosity of $\sim~10^{43}-10^{44}~$\ergs could explain the measured outflow mass loading factors ($\approx 0.1$), mass outflow rates ($\approx 20$~\Msunyr), and kinetic powers ($\dot E_{out}\approx 10^{42}$~\ergs), according to theoretical predictions. 
 \end{itemize}

The NIRSpec observations allowed us to connect the 100~pc scale cold molecular outflows revealed by ALMA (e.g., \citealt{Wheeler2020}) with the greater than $1~$kpc scale ionized and neutral outflows identified in the MUSE data cube (e.g., \citealt{Perna2020}). 
These findings highlight the significant role that individual outflows launched by each nucleus during a merger could play in the starburst-quasar evolutionary sequence. These outflows propagate in multiple directions on kiloparsec scales, potentially impacting the overall physical and  dynamical conditions of the ISM in the host galaxy and are at odds with the more compact, spherical outflows detected in other nearby systems (e.g., \cite{Venturi2021}). 
Simulations predict that multidirectional outflows could lead to more uniform quenching of SF throughout the galaxy. In contrast, collimated outflows 
may be limited to specific directions and may allow SF to continue in other parts of the galaxy (see, e.g., \citealt{Wagner2012, Nelson2019, Torrey2020, Sivasankaran2024}).  

We also investigated the properties of the ISM not participating in the outflow in terms of recent SF traced by recombination lines and obscuration. Additionally, we compared the stellar and the gas kinematics in both the close surroundings of the two nuclei and across the large-scale disk. The key findings of this analysis are summarized below. 

\begin{itemize}

\item 
Consistent with previous works, we found that SFR calculated with recombination lines ($\sim$ 8~\Msunyr) differs from the SF computed with FIR and radio ($\sim~230$~\Msunyr). This result possibly indicates that a large fraction of the SF is completely obscured at near-IR wavelengths. Alternatively, the FIR measurements of SFR may trace SF on different timescales.
\item 
We estimated the attenuation from different emission line ratios: \feii~1.644~$\mu$m/\feii~1.257~$\mu$m, Br$\gamma$/Pa$\beta$, and Pa$\alpha$/Pa$\beta$. These values are consistent within 20\%. The highest values are found at the position of the E nucleus, A$_V \sim $10 mag, and the lowest values (A$_V \sim~2-4$~mag) are in the outer regions ($\sim~400$~pc distance from the nuclei).
\item 
The stellar and gas kinematics show a relatively good match, with the stellar component following the ionized gas rotation and appearing to rotate faster than the hot molecular gas.

 \end{itemize}
The near-IR, spatially resolved spectroscopic data of Arp 220 presented in this work demonstrates the power of JWST/NIRSpec to disentangle the complex kinematics in the nuclear regions and to understand the interplay between multiphase outflows in the circumnuclear environments of nearby ULIRGs. Moreover, the data shed light on the feedback mechanisms in this peculiar phase of galaxy evolution.

\section{Data availability}
The data cube, the spectra shown in Fig. \ref{Integrated Spectra}, and line maps presented in Fig. \ref{Moments} are only available in electronic form at the CDS via anonymous ftp to \url{cdsarc.u-strasbg.fr} (130.79.128.5) or via \url{http://cdsweb.u-strasbg.fr/cgi-bin/qcat?J/A+A/}

 \begin{acknowledgements}
 This article was produced while attending the PhD program in Space Science and Technology at the University of Trento, Cycle XXXVIII, with the support of a scholarship financed by the Ministerial Decree no. 351 of 9th April 2022, based on the NRRP - funded by the European Union - NextGenerationEU - Mission 4 "Education and Research", Component 1 "Enhancement of the offer of educational services: from nurseries to universities” - Investment 4.1 “Extension of the number of research doctorates and innovative doctorates for public administration and cultural heritage" - E63C22001340001. 
MP, SA, and BRP acknowledge support from the research project PID2021-127718NB-I00 of the Spanish Ministry of Science and Innovation/State Agency of Research (MCIN/AEI/10.13039/501100011033).
IL acknowledges support from PID2022-140483NB-C22 funded by AEI 10.13039/501100011033 and BDC 20221289 funded by MCIN by the Recovery, Transformation and Resilience Plan from the Spanish State, and by NextGenerationEU from the European Union through the Recovery and Resilience Facility.
GC, AM acknowledge support from PRIN-MUR project "PROMETEUS" (202223XPZM), the INAF Large Grant 2022 “The metal circle: a new sharp view of the baryon cycle up to Cosmic Dawn with the latest generation IFU facilities” and INAF Large Grant 2022 "Dual and binary SMBH in the multi-messenger era".  
KF acknowledges support through the ESA research fellowship programme.
AJB acknowledges funding from the “FirstGalaxies” Advanced Grant from the European Research Council (ERC) under the European Union’s Horizon 2020 research and innovation program (Grant agreement No. 789056).
FDE and RM acknowledge support by the Science and Technology Facilities Council (STFC), by the ERC through Advanced Grant 695671 ``QUENCH'', and by the
UKRI Frontier Research grant RISEandFALL. RM also acknowledges funding from a research professorship from the Royal Society.
MPS acknowledges support from grants RYC2021-033094-I and CNS2023-145506 funded by MICIU/AEI/10.13039/501100011033 and the European Union NextGenerationEU/PRTR.
 \end{acknowledgements}

\bibliographystyle{aa}
\bibliography{references} 



\appendix
\section{G140H stellar kinematics}

\begin{figure}[h!]
    \centering
    \includegraphics[width = 0.5 \textwidth]{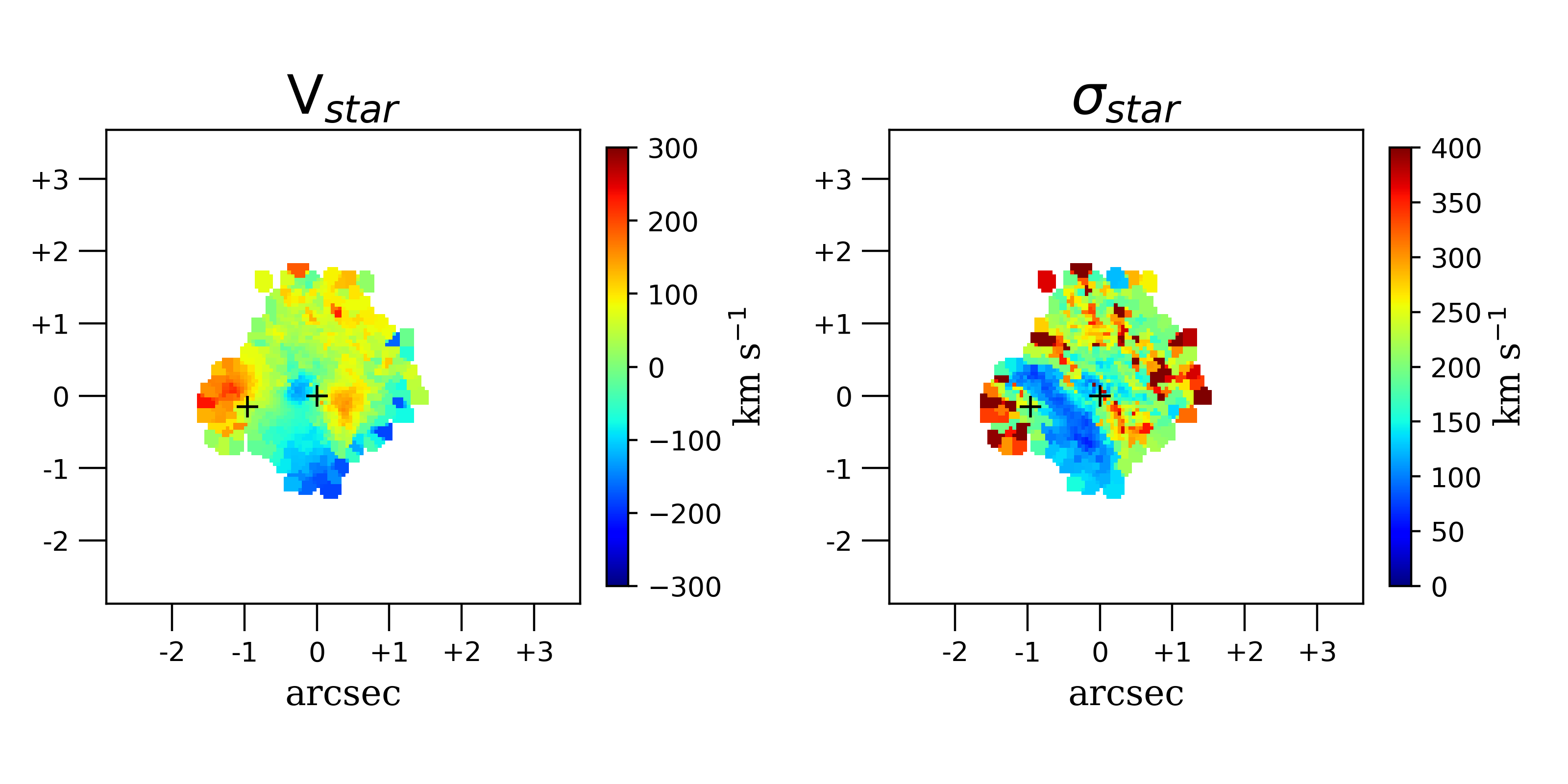}
    \caption{Stellar kinematics derived from the analysis of the first cube (grating G140H/F100LP).}
    \label{1cube_stellar}
\end{figure}

\section{\MOKA}
In the Appendix B, we show the comparison of the data and the kinematic model \MOKA for the SE outflow, HG outflow and NW bubble as discussed in Sect. \ref{sec:modellingwithMOKA}.
\begin{figure*}[b!]
    \centering
    \includegraphics[width = 0.9 \textwidth]{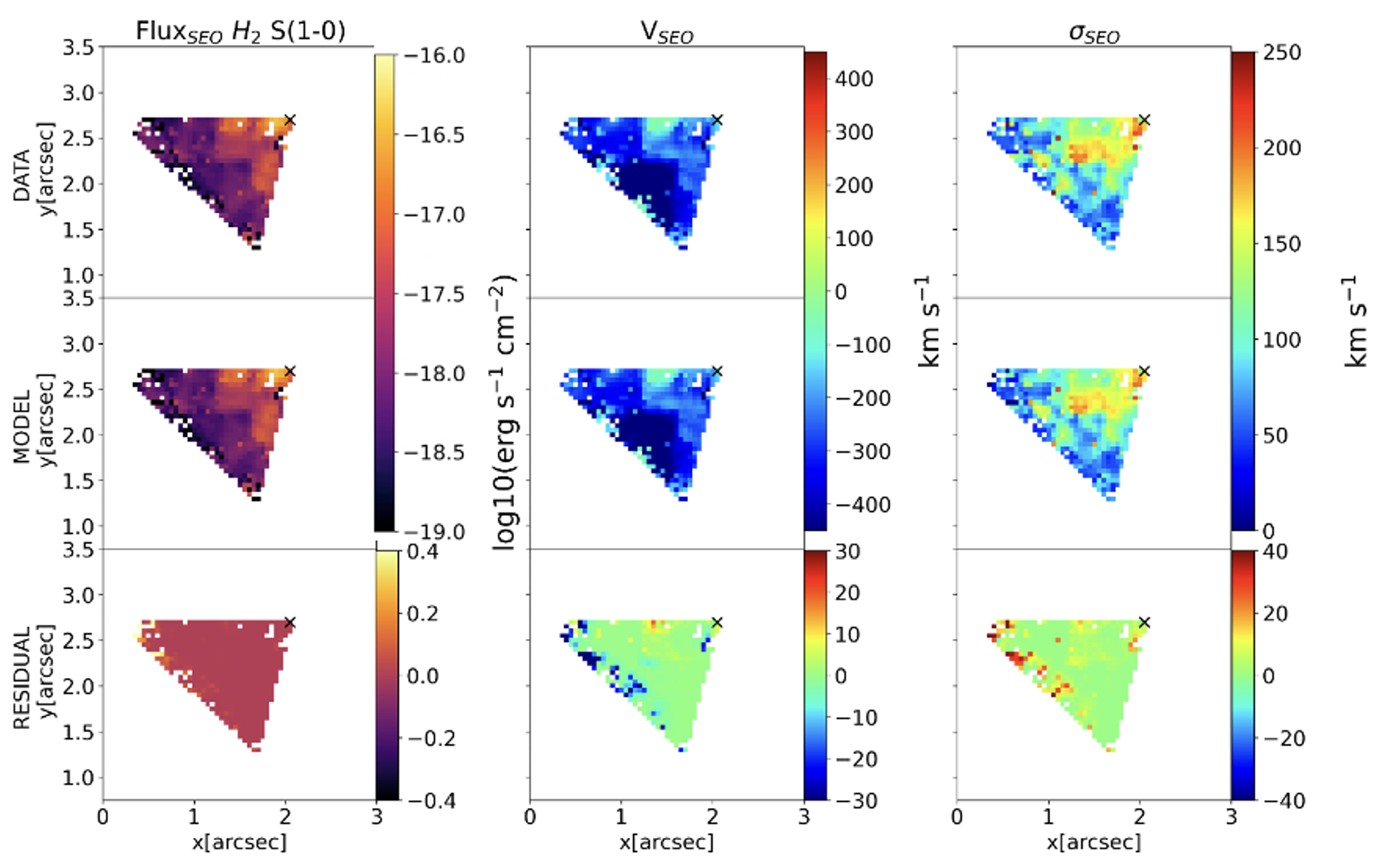}
    \caption{ Comparison between the flux and velocity of H$_{2}$ (top panels) and \MOKA weighted model (middle panels) for the SE conical outflow. The bottom panels present the residuals obtained subtracting the model from the data. }
    \label{SEMokafig}
\end{figure*}

\begin{figure*}[h!]
    \centering
    \includegraphics[width = 0.9 \textwidth]{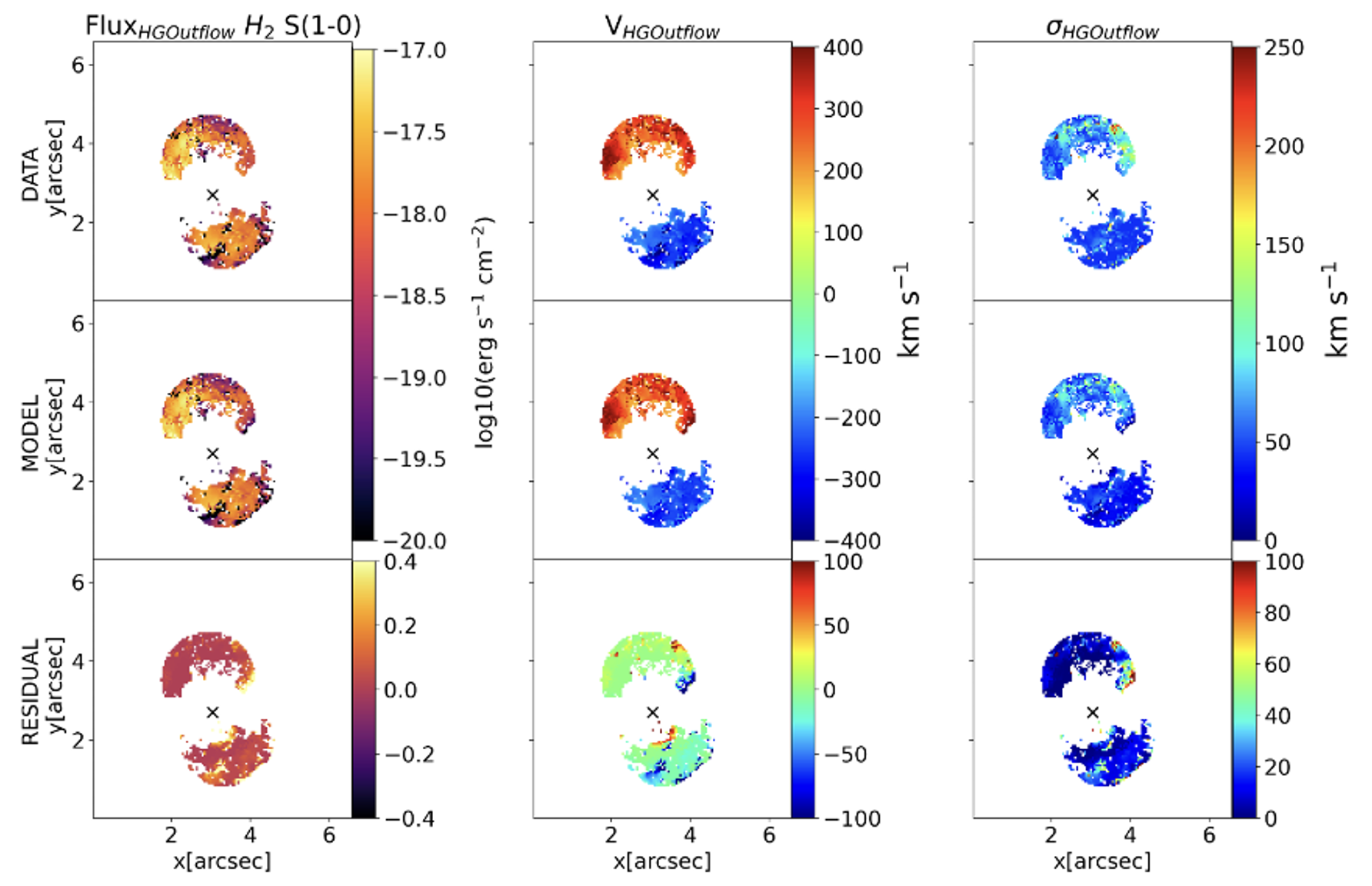}
    \caption{ Same as \ref{SEMokafig} for the HG outflow. }
    \label{Mokafig}
\end{figure*}

\begin{figure*}[h!]
    \centering
    \includegraphics[width = 0.9 \textwidth]{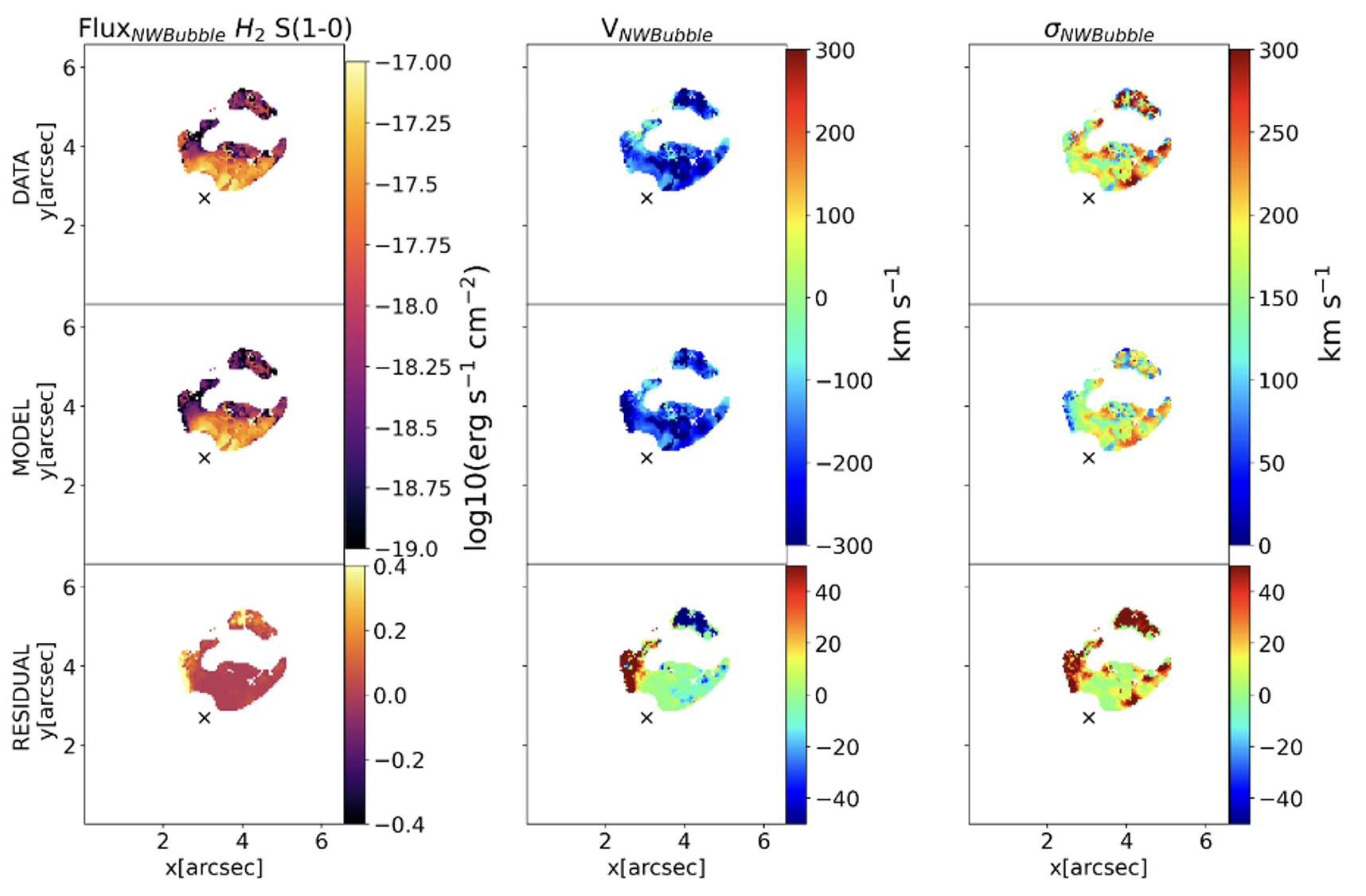}
    \caption{ Same as \ref{SEMokafig} for the NW bubble. }
    \label{Mokafig}
\end{figure*}


\label{lastpage}
\end{document}